\RequirePackage{fix-cm}
\documentclass[smallextended]{svjour3-mod}   
\smartqed  
\usepackage{amsmath,amssymb,latexsym,stmaryrd}
\usepackage{bussproofs}
\usepackage{url}
\usepackage{graphicx}
\usepackage{soul,color}
\usepackage{amsmath}
\newcommand{\foldn}{\ensuremath{FO\lambda^{\Delta I\!\!N}}}
\usepackage{cmll}

\usepackage{amssymb}
\usepackage{url}

\usepackage{proof}

\newcommand{\encode}[2]{\varepsilon_{#1}{(#2)}}
\newcommand{\decode}[2]{\delta_{#1}{(#2)}}

\begin{document}

\title{Formalization of Metatheory of the Quipper Quantum
 Programming Language in a Linear Logic}

\titlerunning{Formalization the Quipper Metatheory}        

\author{Mohamed Yousri Mahmoud \and Amy P. Felty}

\authorrunning{M. Y. Mahmoud\and
 A. P. Felty} 

\institute{
              School of Electrical Engineering and Computer Science,
             University of Ottawa, Ottawa, Canada\\
              \email{myousri@uottawa.ca, afelty@uottawa.ca}}           


\date{December 10, 2018}

\maketitle

\begin{abstract}
We develop a linear logical framework within the
  Hybrid system and use it to reason about the type system of a
  quantum lambda calculus.  In particular, we consider a practical
  version of the calculus called Proto-Quipper, which contains the
  core of Quipper. Quipper is a new quantum programming language under
  active development and recently has gained much popularity among the
  quantum computing communities.  Hybrid is a system that is designed
  to support the use of higher-order abstract syntax (HOAS) for
  representing and reasoning about formal systems implemented in the
  Coq Proof Assistant.  In this work, we extend the system with a
  linear specification logic (SL) in order to reason about the linear
  type system of Quipper.  To this end, we formalize the semantics of
  Proto-Quipper by encoding the typing and evaluation rules in the SL,
  and prove type soundness.
\end{abstract}

\section{Introduction}
Quipper is a functional programming language designed for implementing
quantum algorithms~\cite{GreenEtAl:PLDI12}.  The mathematical
foundations of the Proto-Quipper fragment, which retains much of the
important expressive power of the full language, is developed
in~\cite{RossPhD15}.  As the authors themselves have
noted~\cite{SelingerEmail16}, there is a great deal of subtlety in the
definitions of the syntax and semantics, and many details were
fine-tuned during the process of proving the type soundness result.
This process certainly would have benefited from formalization and the
ability to recheck proofs after each change in the definitions.  Quipper
is a relatively new language, and additional metatheory will be proved
as it is developed.  Providing an environment within a proof assistant
that allows the developers to simultaneously develop and formalize
this metatheory is a central goal of our work.

The Hybrid system~\cite{FeltyMomigliano:JAR12} provides support for
reasoning about \emph{object languages} (OLs) such as programming
languages and other formal systems using \emph{higher-order abstract
  syntax} (HOAS), sometimes referred to as ``lambda-tree
syntax'' \cite{MillerPalamidessi:ACMSurveys99,PfenningElliot:PLDI88}.
It is implemented as a two-level system, an approach first introduced
in the $\foldn$ logic \cite{McDowellMiller:TOCL01}.  Using this
approach, the specification of the semantics of the OL and the
meta-level reasoning about it are done within a single system but at
different levels.  In the case of Hybrid, an intermediate level is
introduced by inductively defining a \emph{specification logic} (SL)
in Coq, and OL judgments are encoded in the SL.
The version of Hybrid we use here is implemented as two distinct
libraries in the Coq Proof Assistant.  The first we adopt without
change, while the development of a new version of the second is part
of the contributions of this work.

The first Hybrid library provides support for
expressing the syntax of OLs.  Using HOAS, binding constructs of the
OL are encoded using lambda abstraction in Coq.
For instance, for Proto-Quipper, type $\mathtt{qexp}$ will represent
terms or programs, and $\texttt{App}$ of type
$\mathtt{qexp}\rightarrow\mathtt{qexp}\rightarrow\mathtt{qexp}$ and
$\texttt{Fun}$ of type
$(\underline{\mathtt{qexp}}\rightarrow\mathtt{qexp})\rightarrow\mathtt{qexp}$
will represent application and abstraction, respectively, of the
linear lambda calculus, which forms the core of the
Proto-Quipper language.  Lambda abstraction in Proto-Quipper is a
\emph{binder} because the name of a variable is bound in the body of
the abstraction.  For example, the term $\lambda x . \lambda y . x 
y$ can be encoded as $(\mathtt{Fun}~(\lambda x.\mathtt{Fun}~(\lambda
y.(\mathtt{App}~x~y)))$. Note that $\mathtt{qexp}$ cannot be defined
inductively because of the (underlined) negative occurrence of
$\mathtt{qexp}$ in the type of $\mathtt{Fun}$.
Hybrid provides an underlying low-level de Bruijn representation of
terms to which the HOAS representation can be mapped internally; the
user works directly with the higher level HOAS representation.

Using such a representation, $\alpha$-conversion at the meta-level
directly represents bound variable renaming at the object-level, and
meta-level $\beta$-conversion can be used to directly implement
object-level substitution.  As a consequence, we avoid the need to
develop large libraries of lemmas devoted to operations dealing with
variables, such as capture-avoiding substitution, renaming, and fresh
name generation.
In proof developments that use a first-order syntax, such libraries
are often significantly larger than the formalization of the main
metatheory results.

The second Hybrid library defines the SL and provides support for
encoding OL judgments and inference rules, and for reasoning about
them.
The need for different levels arises in Hybrid because there are OL
judgments that cannot be encoded as inductive propositions in Coq. As an
example, we consider assigning simple types to lambda terms.
The standard rule is as follows:
$$\AxiomC{$\Gamma,x:T \vdash t:T'$}
\RightLabel{}
\UnaryInfC{$\Gamma \vdash \lambda x.t : T \rightarrow T'$}
\DisplayProof$$
Let \texttt{qtp} be the type of OL types, and let \texttt{arr} be a
constant of type $\mathtt{qtp} \rightarrow \mathtt{qtp} \rightarrow
\mathtt{qtp}$ for constructing arrow types.  Let \texttt{typeof} be a
predicate expressing the relation between a term and its type.  If we
consider the
HOAS encoding of the above rule, using for example, the techniques
introduced in the logical framework LF~\cite{Harper93jacm}, we can
encode the above rule as the following formula:
$$\begin{array}{l}
\forall T,T':\mathtt{qtp}.\;\forall t:\mathtt{qexp} \rightarrow
    \mathtt{qexp}. \\
\quad (\forall x : \mathtt{qexp}.\; \underline{\mathtt{typeof}~x~T}
\rightarrow \mathtt{typeof}~(t~x)~T') \rightarrow {} \\
\quad\quad{}\mathtt{typeof}~(\mathtt{Fun}~t)~(\mathtt{arr} \; T \; T')
\end{array}$$
The second line contains a formula with embedded implication and
universal quantification (known as \emph{hypothetical} and
\emph{parametric} judgments), in particular universal quantification
over variable $x$ of type $\mathtt{qexp}$ and an internal assumption
(underlined) about the type of this $x$.  We note that the
\texttt{typeof} predicate cannot be expressed inductively because this
underlined occurrence is negative.  A two-level system solves this
problem by encoding OL predicates inside the SL, where negative
occurrences are allowed.  We will see how this is done in
Section~\ref{sec:pqsyntax}, where we encode rules such as the typing
rule for lambda abstraction in the quantum lambda calculus.

The linear SL we implement here is an extension of the ordered linear
SL implemented in~\cite{FeltyMomigliano:JAR12} and the ordered and
linear SLs presented in~\cite{MartinPhD2010}.  We extend and adapt
this previous work to the much larger case study considered here, and
we design the new SL to be general so that it can be adopted directly
for reasoning about a variety of other OLs with linear features.

We formalize the key property of \emph{type soundness} (also called
\emph{subject reduction}) of Proto-Quipper, which requires several
important lemmas about context subtyping.

The issue of the \emph{adequacy} of HOAS representations is important
(see~\cite{Harper93jacm}), which here means that we must prove that
the encoding in Hybrid does indeed represent the intended language.
The work described here includes extending previous work on
adequacy to our setting, where both the OL (Proto-Quipper) and the SL
(a linear logic) are more complex than those considered previously.

We begin in the next section with background material on
Proto-Quipper.  Then in Section~\ref{sec:syntax}, we present the
encoding of types and of the subtyping relation in Coq.  Proto-Quipper
types and subtyping can be encoded directly using inductive types.
There are no binders and no substitution.  One of the strengths of
Hybrid as implemented in Coq is that it is straightforward to combine
such direct encodings of OL syntax with encodings that use the HOAS
and SL facilities provided by Hybrid.
In Section~\ref{sec:hybrid}, we present the Hybrid system including
both background information on the first Coq library, as well as our
new implementation of the second library, which encodes our linear SL
in Coq and develops some important general meta-theoretic properties
that help in reasoning about OLs.

The encoding of Proto-Quipper terms and of the typing
rules, along with some properties about them appear in
Section~\ref{sec:encoding}.
The encoding of reduction rules is presented in
Section~\ref{sec:sr}, along with the statement of type soundness and a
discussion of its formal proof.

We discuss adequacy of our encoding
in Section~\ref{sec:adequacy}, and finally, we conclude and
discuss related and future work in Section~\ref{sec:concl}.

The files of our formalization are publicly available~\cite{ourformal}.

\section{Proto-Quipper}
\label{sec:pq}

We give a brief background of the Proto-Quipper language,
focusing on the aspects that are required for understanding the
formalization in later sections.  Proto-Quipper is based on the
quantum lambda calculus and focuses on Quipper's abilities to generate
and manipulate quantum circuits~\cite{RossPhD15}.
The types and terms of Proto-Quipper are defined by the following
grammars:
$$\begin{array}{ccl}
T,U & {}::={} & \mathbf{qubit} \mid
                       1 \mid
                       T \otimes U\\
A,B & {}::={} & \mathbf{qubit} \mid 
              1 \mid \mathop{!}1 \mid
              \mathbf{bool} \mid \mathop{!}\mathbf{bool} \mid
              A \otimes B \mid \mathop{!}(A \otimes B) \mid \\
&&            A \multimap B \mid \mathop{!}(A \multimap B) \mid
              \mathrm{Circ}(T,U) \mid \mathop{!}(\mathrm{Circ}(T,U)) \\
t & {}::={} & q \mid \mathop{*} \mid \langle t_1,t_2 \rangle \\
a,b,c & {}::={} & x \mid q \mid (t,C,a) \mid \mathtt{True} \mid
              \mathtt{False} \mid \langle a, b \rangle \mid
              \mathop{*} \mid
              ab \mid \lambda x.a \mid \\
&&              \mathit{rev} \mid \mathit{unbox} \mid \mathit{box}^T
              \mid
              \mathtt{if}~a~\mathtt{then}~b~\mathtt{else}~c \mid\\
&&            \mathtt{let}\mathop{*}= a~\mathtt{in}~b \mid
              \mathtt{let}~\langle x, y \rangle = a~\mathtt{in}~b
\end{array}$$
Proto-Quipper distinguishes between \emph{quantum data types} ($T,U$) and
\emph{types} ($A,B$) where the former is a subset of the latter, and
similary between \emph{quantum data terms} ($t$) and
\emph{terms} ($a,b,c$). Here $x$ and $y$ are \emph{term variables} from a set
$\cal{V}$, $q$ is a \emph{quantum variable} from a set $\cal{Q}$, and
$C$ is a \emph{circuit constant} from a set $\cal{C}$.  The sets
$\cal{V}$, $\cal{Q}$, and $\cal{C}$ are all assumed to be countably
infinite.

Most types and term constructs come directly from the quantum lambda
calculus, e.g., \cite{SelingerValiron:MSCS06}.  The type
$\mathrm{Circ}(T_1,T_2)$ represents the set of all circuits having an
input interface of type $T_1$ and an output interface of type $T_2$.
A circuit constant $C$ represents a low-level quantum circuit, and a
term $(t,C,a)$ represents a circuit as Proto-Quipper data, where $t$
is a structure representing a finite set of inputs to $C$, and
similarly $a$ represents a finite set of outputs.  In Proto-Quipper,
it is assumed that two functions exists, $In$ and $Out$, from
$\cal{C}$ to the set of all subsets of $\cal{Q}$, where $In(C)$ and
$Out(C)$ are a superset of the set of input and output quantum
variables, respectively, of circuit $C$.  The terms $\mathit{rev}$,
$\mathit{unbox}$, and $\mathit{box}^T$ represent functions on
circuits.  We refer the reader to~\cite{RossPhD15} for further
description of these types and terms.

In quantum computing, variable cloning is prohibited. This features is
reflected in Proto-Quipper by using the modal operator $\mathop{!}$,
where variables having type $!A$ are called duplicable and can be cloned
whereas variables of types that do not follow this pattern are called
linear and cannot be cloned or copied.  Note that instead of
introducing a general $\mathop{!}$ operator on types, we restrict it
to prohibit more than one consecutive occurrence of the bang operator $\mathop{!}$.
This presentation of types differs from the one
in~\cite{RossPhD15}.\footnote{In general, when we stray from the
  original presentation, our intention is to simplify formalization,
  and we only do so when there is a clear equivalence to the original.
  In this case, we simplify the formalization of the subtyping
  relation without changing the semantics of types.  As we will
  discuss in the next section, making this kind of change also led to
  the discovery of a small mistake in the original presentation.}

The use of the bang
operator $\mathop{!}$ introduces a subtyping relation among types: A
variable of type $!A$ obviously is also of type $A$. The subtyping
rules of Proto-Quipper are shown in Figure~\ref{fig:subtyperules}.
The subtyping relation $\mathop{<:}$ is the smallest relation on types
satisfying these rules.
\begin{figure}
\centering
\begin{tabular}{c}
\AxiomC{}
\RightLabel{}
\UnaryInfC{$qubit\mathop{<:}qubit$}
\DisplayProof
\hspace{1cm}
\AxiomC{}
\RightLabel{}
\UnaryInfC{$1\mathop{<:}1$}
\DisplayProof
\hspace{1cm}
\AxiomC{}
\RightLabel{}
\UnaryInfC{$bool\mathop{<:}bool$}
\DisplayProof
\\ \\
\AxiomC{$A_1\mathop{<:}B_1$}
\AxiomC{$A_2\mathop{<:}B_2$}
\RightLabel{}
\BinaryInfC{$A_1\otimes A_2 \mathop{<:} B_1\otimes B_2$}
\DisplayProof
\hspace{1cm}
\AxiomC{$A_2\mathop{<:}A_1$}
\AxiomC{$B_1\mathop{<:}B_2$}
\RightLabel{}
\BinaryInfC{$A_1\multimap B_1 \mathop{<:} A_2\multimap B_2$}
\DisplayProof\\ \\
\AxiomC{$T_2\mathop{<:}T_1$}
\AxiomC{$U_1\mathop{<:}U_2$}
\RightLabel{}
\BinaryInfC{$Circ(T_1,U_1) \mathop{<:} Circ(T_2,U_2)$}
\DisplayProof
\hspace{1cm}
\AxiomC{$A\mathop{<:}B$}
\RightLabel{$\dagger$}
\UnaryInfC{$\mathop{!}A\mathop{<:} B$}
\DisplayProof
\hspace{1cm}
\AxiomC{$A\mathop{<:}B$}
\RightLabel{$\dagger$}
\UnaryInfC{$\mathop{!}A\mathop{<:} \mathop{!}B$}
\DisplayProof
\end{tabular}

\bigskip
$\dagger$ Note that the last two rules have the proviso that $A$ and
$B$ do not have a leading $!$.
\caption{Subtyping rules for Proto-Quipper \label{fig:subtyperules}}
\end{figure}

The Proto-Quipper typing judgment has the form $\Phi;Q\vdash a:A$.
In this sequent, $\Phi$ is a finite set of typing declarations of
the form $x:A$ where $x$ is a variable and $A$ is a type ($A$ may have
the form $!C$ or not).  In the presentation of the rules, $\Phi$
always appears in two parts $\Phi',\mathop{!}\Psi$ where the types
in the latter all follow the pattern $\mathop{!}A$, while those in
the former never contain a leading $\mathop{!}$. $Q$ is a quantum
context containing a finite set of quantum variables, typically the
free quantum variables in $a$.  Also $a$ is a term and
$A$ is a type. The typing rules are shown in Figure~\ref{fig:typingrules}.
\begin{figure}
\centering
\begin{tabular}{c}
\AxiomC{}
\AxiomC{}
\RightLabel{$ax_q$}
\BinaryInfC{$\mathop{!}\Psi;{q}\vdash q:\mathbf{qubit}$}
\DisplayProof
\hspace{1cm}
\AxiomC{}
\AxiomC{}
\RightLabel{$ax_x$}
\BinaryInfC{$\mathop{!}\Psi,x:A;\cdot\vdash x:A$}
\DisplayProof\\ \\

\AxiomC{$\mathop{!}\Psi;\cdot\vdash a:\mathop{!}^n A$}
\AxiomC{$\mathop{!}^n A \mathop{<:} B$}
\RightLabel{$ax_c$}
\BinaryInfC{$\mathop{!}\Psi;\cdot\vdash a:B$}
\DisplayProof
\hspace{1cm}
\AxiomC{$\mathop{!}A_c(T,U) \mathop{<:} B$}
\RightLabel{$\mathit{cst}$}
\UnaryInfC{$\mathop{!}\Psi;\cdot\vdash c:B$}
\DisplayProof\\ \\

\AxiomC{}
\AxiomC{}
\RightLabel{$\ast_{i}$}
\BinaryInfC{$\mathop{!}\Psi;\cdot\vdash \ast:\mathop{!}^n 1$}
\DisplayProof
\hspace{.5cm}
\AxiomC{}
\AxiomC{}
\RightLabel{$\top$}
\BinaryInfC{$\mathop{!}\Psi;\cdot\vdash \mathtt{True}: \mathop{!}^n bool$}
\DisplayProof
\hspace{.5cm}
\AxiomC{}
\AxiomC{}
\RightLabel{$\bot$}
\BinaryInfC{$\mathop{!}\Psi;\cdot\vdash \mathtt{False}:\mathop{!}^n bool$}
\DisplayProof\\ \\

\AxiomC{$\Phi,x:\mathop{!}^n A;Q\vdash b:B$}
\RightLabel{$\lambda_1$}
\UnaryInfC{$\Phi;Q\vdash \lambda x.b:\mathop{!}^n A\multimap B$}
\DisplayProof
\hspace{1cm}
\AxiomC{$\mathop{!}\Psi,x:\mathop{!}^n A;\cdot\vdash b:B$}
\RightLabel{$\lambda_2$}
\UnaryInfC{$\mathop{!}\Psi;\cdot\vdash \lambda x.b:\mathop{!}(\mathop{!}^n A\multimap B)$}
\DisplayProof\\ \\

\AxiomC{$\Phi_1,\mathop{!}\Psi;Q_1\vdash c:A\multimap B$}
\AxiomC{$\Phi_2,\mathop{!}\Psi;Q_2\vdash a:A$}
\RightLabel{$\mathit{app}$}
\BinaryInfC{$\Phi_1,\Phi_2,\mathop{!}\Psi;Q_1,Q_2\vdash ca:B$}
\DisplayProof\\ \\

\AxiomC{$\Phi_1,\mathop{!}\Psi;Q_1\vdash a:\mathop{!}^n A$}
\AxiomC{$\Phi_2,\mathop{!}\Psi;Q_2\vdash b:\mathop{!}^n B$}
\RightLabel{$\otimes_{i}$}
\BinaryInfC{$\Phi_1,\Phi_2,\mathop{!}\Psi;Q_1,Q_2\vdash \langle a,b\rangle:\mathop{!}^n(A\otimes B)$}
\DisplayProof\\ \\

\AxiomC{$\Phi_1,\mathop{!}\Psi;Q_1\vdash b:\mathop{!}^n(B_1\otimes B_2)$}
\AxiomC{$\Phi_2,\mathop{!}\Psi,x:\mathop{!}^n B_1,y:\mathop{!}^n B_2;Q_2\vdash a:A$}
\RightLabel{$\otimes_{e}$}
\BinaryInfC{$\Phi_1,\Phi_2,\mathop{!}\Psi;Q_1,Q_2\vdash \mathtt{let}\ \langle x,y\rangle = b \ \mathtt{in}\ a: A)$}
\DisplayProof\\ \\

\AxiomC{$\Phi_1,\mathop{!}\Psi;Q_1\vdash b:\mathop{!}^n 1$}
\AxiomC{$\Phi_2,\mathop{!}\Psi;Q_2\vdash a: A$}
\RightLabel{$\ast_e$}
\BinaryInfC{$\Phi_1,\Phi_2,\mathop{!}\Psi;Q_1,Q_2\vdash \mathtt{let}\ \ast = b \ \mathtt{in}\ a: A$}
\DisplayProof\\ \\

\AxiomC{$\Phi_1,\mathop{!}\Psi;Q_1\vdash b:bool$
  \hspace{.5cm}
  $\Phi_2,\mathop{!}\Psi;Q_2\vdash a_1:A$
  \hspace{.5cm}
  $\Phi_2,\mathop{!}\Psi;Q_2\vdash a_2:A$}
\RightLabel{$\mathit{if}$}
\UnaryInfC{$\Phi_1,\Phi_2,\mathop{!}\Psi;Q_1,Q_2\vdash \mathtt{if}\ b\ \mathtt{then}\ a_1\ \mathtt{else}\ a_2:A$}
\DisplayProof\\ \\

\AxiomC{$Q_1\vdash t:T$
  \hspace{.5cm}
  $\mathop{!}\Psi;Q_2\vdash a:U$
  \hspace{.5cm}
  $In(C) = Q_1$
  \hspace{.5cm}
  $Out(C)=Q_2$}
\RightLabel{$\mathit{circ}$}
\UnaryInfC{$\mathop{!}\Psi;\cdot\vdash (t,C,a):\mathop{!}^n \mathtt{Circ}(T,U)$}
\DisplayProof\\ \\
\end{tabular}
\caption{Typing rules for Proto-Quipper \label{fig:typingrules}}
\end{figure}
We write $\cdot$ to represent an empty context.  The ``;" is used to
separate the typing context from the quantum context whereas the ``,"
is used to append two contexts.  The rules containing $\mathop{!}^n$
abbreviate two distinct rules, one where $n=0$, i.e., there is no
leading $\mathop{!}$, and one where $n=1$.
These rules are taken from~\cite{RossPhD15}, with some modifications
that do not change the semantics.  For example, when $\mathop{!}^n$
appears in the rules there, $n$ can be any natural number.  The
restriction here takes into account our modifcation to the syntax of
types discussed above.  Also, the $ax_x$ and $ax_c$ rules replace a
single rule for subtyping in~\cite{RossPhD15}.  Here, we give an
initial rule for term variables ($ax_x$), similar to the one for
quantum variables ($ax_q$).  When using LF-style hypothetical and
parametric judgments, such rules for variables are implicit; they do
not appear in the encoding of the inference rules in the SL.  In
general, we avoid explicit treatment of variables whenever possible to
get the full advantage of HOAS encodings.  In addition, the subtyping
rule in~\cite{RossPhD15} is restricted only to term variables.  Again
to avoid specific reasoning about variables, our $ax_c$ rule is valid
for any valid expression of Proto-Quipper.

In the $\mathit{(cst)}$ rule $c$ ranges over the set
$\{\mathit{box}, \mathit{unbox}, \mathit{rev}\}$, and we write
$\mathit{box}(T,U)$ as $\mathit{box}^T(U)$ to make explicit that the
constant $\mathit{box}$ is always annotated with its type (see the
grammar).

$A_\mathit{{box}^T}$, $A_\mathit{unbox}$, and $A_\mathit{rev}$ are
defined as follows:
$$\begin{array}{rcl}
  A_\mathit{{box}^T}(U) & := & \mathop{!}(T\multimap U)\multimap
  \mathop{!}\mathrm{Circ}(T,U) \\
  A_\mathit{unbox}(T,U) & := & \mathrm{Circ}(T,U)\multimap
  \mathop{!}(T\multimap U)\\
  A_\mathit{rev}(T,U) & := & \mathrm{Circ}(T,U)\multimap
  \mathop{!}\mathrm{Circ}(U,T)
\end{array}$$

\section{Proto-Quipper Types}
\label{sec:syntax}

In the previous section, the types supported in Proto-Quipper were
presented using a context free grammar, where there are two classes of
types. Here, we consider a single ``universal'' class of types, and
then define predicates that discriminate between the two classes.  We
start the formalization of the Proto-Quipper types (file
ProtoQuipperTypes.v) by defining an inductive type of the universal
class:
\begin{flushleft}
\texttt{Inductive~qtp:~Type~:=}\\
\texttt{~~qubit:~qtp}
\texttt{|~one:~qtp}
\texttt{|~bool:~qtp}\\
\texttt{|~tensor:~qtp~-{>}~qtp~-{>}~qtp}
\texttt{|~arrow:~qtp~-{>}~qtp~-{>}~qtp}\\
\texttt{|~circ:~qtp~-{>}~qtp~-{>}~qtp}
\texttt{|~bang:~qtp~-{>}~qtp.}\\
\end{flushleft}
Certainly, the above definition does not  model Proto-Quipper
types. For instance, it does allow for the  parameters of the circuit
type constructor to be of the general class, whereas the argument
expressions are supposed to be of the quantum data type. Accordingly, we
define the inductive predicate \texttt{valid} that captures the notion
of quantum data types (grammar $T,U$):
\begin{flushleft}
\texttt{Inductive~valid:~qtp~-{>}~Prop~:=}\\
\texttt{~~Qubit:~valid~qubit}\\
\texttt{|~One:~valid~one~}\\
\texttt{|~Tensor:~forall~A1~A2,~valid~A1~-{>}~valid~A2~-{>}}\\
\texttt{~~valid~(tensor~A1~A2).}\\
\end{flushleft}
Then, we define the general \texttt{validT} predicate to
identify the general types (grammar $A,B$):
\begin{flushleft}
\texttt{Inductive~validT:~qtp~-{>}~Prop~:=}\\
\texttt{~~vQubit:~validT~qubit}\\
\texttt{|~bQubit:~validT~(bang~qubit)}\\
$\vdots$\\
\texttt{|~vTensor:~forall~A~B,~validT~A~-{>}~validT~B~-{>}}\\
\texttt{~~validT~(tensor~A~B)~}\\
\texttt{|~bTensor:~forall~A~B,~validT~A~-{>}~validT~B~-{>}~}\\
\texttt{~~validT~(bang~(tensor~A~B))~}\\
$\vdots$\\
\texttt{|~vCirc:~forall~A~B,~valid~A~-{>}~valid~B~-{>}}\\
\texttt{~~validT~(circ~A~B)}\\
\texttt{|~bCirc:~forall~A~B,~valid~A~-{>}~valid~B~-{>}~}\\
\texttt{~~validT~(bang~(circ~A~B)).}\\
\end{flushleft}
In longer definitions such as this one, we often omit parts when it is
clear how to fill in the complete idea of the definition.
We have proved that \texttt{(valid A)} implies \texttt{(validT A)},
confirming that class $T,U$ is a subclass of $A,B$.

The last step in the formalization of the Proto-Quipper types is the
development of the subtyping relation.  The following inductive
proposition directly encodes the rules of
Figure~\ref{fig:subtyperules}.
\begin{flushleft}
\texttt{Inductive~Subtyping:~qtp~-{>}~qtp~-{>}~Prop~:=}\\
\texttt{~~QubitSub:~Subtyping~qubit~qubit}\\
\texttt{|~OneSub:~Subtyping~one~one~}\\
$\vdots$\\
\texttt{|~CircSub:~forall~A1~A2~B1~B2,~}\\
\texttt{~~Subtyping~A2~A1~-{>}~Subtyping~B1~B2~-{>}}\\
\texttt{~~validT~(circ~A1~B1)~-{>}~validT~(circ~A2~B2)-{>}}\\
\texttt{~~Subtyping~(circ~A1~B1)~(circ~A2~B2)}\\
\texttt{|~BangSub1:~forall~A~B,~Subtyping~A~B~-{>}}\\
\texttt{~~validT~(bang~A)~-{>}~Subtyping~(bang~A)~B}\\
\texttt{|~BangSub2:~forall~A~B,~Subtyping~A~B~-{>}}\\
\texttt{~~validT~(bang~A)~-{>}~Subtyping~(bang~A)~(bang~B).}\\
\end{flushleft}
The main difference between this definition and the original subtyping
rules in~\cite{RossPhD15} are the last two rules. As mentioned in
Section~\ref{sec:pq}, they are stated so that they prevent consecutive
applications of the $\mathop{!}$ operator.  The first of these rules
(\texttt{BangSub1}) is concerned with the weakening of subtype
\texttt{A} by adding the bang operator. The second rule concerns
weakening both sides of the subtyping relation. The use of the
predicate \texttt{validT} throughout the definition ensures that
non-Quipper types are ruled out. This presentation of the subtyping
rules for the bang operator make the formal proof of the transitivity
of the subtyping relation much easier; in particular it avoids a lot
of unneeded induction cases. An important theorem, which states that
the subtyping relation implies the validity of its arguments, is
stated below.
\begin{flushleft}
\texttt{Theorem~SubAreVal:~forall~A~B,~Subtyping~A~B~-{>}~}\\
\texttt{~~validT~A~/{\char'134}~validT~B.}\\
\end{flushleft}

Given the above definition of Proto-Quipper subtyping, we
successfully verified all the required specifications reported in
\cite{RossPhD15}, which ensures that the implemented definition
follows the intended behavior. In the following, we list examples of
the formally proven properties:
\begin{flushleft}
\texttt{Theorem~Subtyping\_Prop1:~forall~B,~}\\
\texttt{~~Subtyping~qubit~B~-{>}~B~=~qubit.}\\
\end{flushleft}
\noindent Similar theorems have been proven for the other base types \texttt{one} and \texttt{bool}.
The following specification ensures that whenever the top-level type constructor of the super-type is the bang operator, then the subtype should be too.
\begin{flushleft}
\texttt{Theorem~Subtyping\_Prop6:~forall~A~B1,}\\
\texttt{~~Subtyping~A~(bang~B1)~-{>}}\\
\texttt{~~exists~A1,~A~=~(bang~A1)~/{\char'134}~Subtyping~A1~B1.}\\
\end{flushleft}
The second conjunct of this theorem can be concluded by inversion
using the \texttt{Bang\_Sub2} rule. It is important to know that this
property cannot be proven for the original subtyping relation reported
in \cite{RossPhD15}. Let us consider the case of \texttt{(Subtyping
  !one !!one)}, from which one cannot conclude that \texttt{(Subtyping
  one !one)}. Thanks to the formal proof, we were able to spot this
ill-formed condition. The author has been contacted and confirmed the
mistake.

Similar to the bang constructor, we prove that if the outermost type constructor of the subtype (not the super-type) is \texttt{arrow} then the super-type should be too:
\begin{flushleft}
\texttt{Theorem~Subtyping\_Prop2:~forall~A1~A2~B,}\\
\texttt{~~Subtyping~(arrow~A1~A2)~B~-{>}}\\
\texttt{~~exists~B1~B2,~}\\
\texttt{~~B~=~arrow~B1~B2~/{\char'134}~Subtyping~B1~A1~/{\char'134}~Subtyping~A2~B2.}\\
\end{flushleft}
\noindent Similar theorems have been developed for the other type constructors \texttt{tensor} and \texttt{circ}. Finally, we provide two important properties of the subtyping relation: reflexivity and transitivity:
\begin{flushleft}
\texttt{Theorem~sub\_ref:~forall~A,~validT~A~-{>}~Subtyping~A~A.}\\
\end{flushleft}
\noindent Note that reflexivity is subject to the validity of
\texttt{A}, i.e., it belongs to the Proto-Quipper types.

\begin{flushleft}
\texttt{Theorem~sub\_trans:~forall~A~B~C,}\\
\texttt{~~Subtyping~A~B~-{>}~Subtyping~B~C~-{>}~Subtyping~A~C.}\\
\end{flushleft}
\noindent Note that transitivity does not require that validity of \texttt{A}, \texttt{B}, and \texttt{C} since it is implicitly imposed from the subtyping antecedents (see \texttt{Theorem SubAreVal}).

This concludes the Proto-Quipper types formalization, where we considered the formal development of valid Proto-Quipper types and the subtyping relation on them.

\section{Two-Level Hybrid}
\label{sec:hybrid}

As mentioned, the purpose of the first Hybrid library (implemented as
Hybrid.v) is to provide support for expressing the higher-order
abstract syntax of OLs.  This file is described in
Section~\ref{sec:hybridhoas}, focusing on the aspects that are
required for understanding the formalization in later sections.  The
second library (implemented as LSL.v) is described in
Section~\ref{sec:sl} and contains the encoding of the specification
logic (SL), which provides the two-level reasoning capabilities as
described earlier.  Hybrid allows the use of different SLs, and as
mentioned, one of the contributions of this paper is to present a new
one.
In fact, our goal is to provide a general framework for reasoning
about a large class of OLs that have linear features, and preliminary
work toward this goal is discussed in~\cite{MahmoudFelty:LSFA17}.
This aspect of our work follows the tradition of a variety of linear
logical frameworks that have been introduced in the literature (though
none that we know of is maintained in a such a way that we could
easily adopt it for our work).  In addition, there is growing body of
OLs with linear features that would benefit from such a framework.
Examples of such frameworks and applications are discussed in
Section~\ref{sec:concl}.

\subsection{Expressing Syntax of Object Languages in Hybrid}
\label{sec:hybridhoas}

At the core is a type
\texttt{expr} that encodes a de Bruijn representation of lambda terms.
It is defined inductively in Coq as follows:
\begin{flushleft}
\texttt{Inductive~expr:~Set~:=}\\
\texttt{|~CON:~con~-{>}~expr}\\
\texttt{|~VAR:~var~-{>}~expr}\\
\texttt{|~BND:~bnd~-{>}~expr}\\
\texttt{|~APP:~expr~-{>}~expr~-{>}~expr}\\
\texttt{|~ABS:~expr~-{>}~expr.}\\
\end{flushleft}
Here, \texttt{VAR} and \texttt{BND} represent bound and free
variables, respectively, and \texttt{var} and \texttt{bnd} are defined
to be the natural numbers.  The type \texttt{con} is a parameter to be
filled in when defining the constants used to represent an OL.  The
library then includes a series of definitions used to define the
operator \texttt{lambda} of type
$(\texttt{expr}\rightarrow\texttt{expr})\rightarrow\texttt{expr}$,
which provides the capability to express OL syntax using HOAS,
including negative occurrences in the types of binders.  Expanding its
definition fully down to primitives gives the low-level de Bruijn
representation, which is hidden from the user when reasoning about
metatheory. In fact, the user only needs \texttt{CON}, \texttt{VAR},
\texttt{APP}, and \texttt{lambda} to define operators for OL syntax.  Two
other predicates from the Hybrid library will appear in the proof
development, $\texttt{proper}:\mathtt{expr}\rightarrow\mathtt{Prop}$
and $\mathtt{abstr}:(\mathtt{expr} \rightarrow \mathtt{expr})
\rightarrow \mathtt{Prop}$.
The $\mathtt{proper}$ predicate rules out terms that have
occurrences of bound variables that do not have a corresponding binder
(\emph{dangling indices}).  The $\mathtt{abstr}$ predicate is applied
to arguments of \texttt{lambda} and rules out functions of type
$(\mathtt{expr}\rightarrow\mathtt{expr})$ that do not encode
object-level syntax, discussed further in Section~\ref{sec:adequacy}.

As mentioned, the type \texttt{con} is actually a parameter in the
Hybrid library.  This will become explicit when presenting Coq
definitions, where we write $(\texttt{expr}~ \texttt{con})$ as the
type used to express OL terms.  The constants for Proto-Quipper will
be introduced later as an inductive type called \texttt{Econ} and the
type \texttt{qexp} mentioned earlier is an abbreviation for
\texttt{(expr Econ)}.

Before describing our SL in the next section, we include here a
summary of the steps of the implementation of any SL.
Formulas of the SL are implemented as an inductive type \texttt{oo}.
This definition introduces constants for the connectives of the SL and
their Coq types.  The rules of the SL are defined as a Coq inductive
proposition, where each clause represents one rule.  This definition
is called \texttt{seq} and has one argument for each of the elements
of a sequent, which includes the context(s) of assumptions and the
conclusion of the sequent.  It may also include a natural number used
to keep track of the height of a deriviation.

The rules of the OL are also defined as an inductive propostion
called \texttt{prog}.  In our case, \texttt{prog} defines the
inference rules for well-formedness of Proto-Quipper terms,
typing of Proto-Quipper terms, and the reduction relation for
evaluating Proto-Quipper terms.  Its definition uses the
capability to express hypothetical and parametric judgments.
In the library file defining the SL, \texttt{prog} is a parameter to
the definition of \texttt{seq}.

When encoding an OL, the file Hybrid.v must be imported since it is
(usually) required to represent the syntax of the OL, while the file
containing the SL must be imported when defining \texttt{prog}.  When
an element of the syntax can be defined directly as an inductive type,
however, there is no need to import any Hybrid files.  This is the
case for the syntax of types of Proto-Quipper as we have seen.  The
terms of Proto-Quipper, however, cannot be defined inductively;
instead we will define a HOAS representation, which uses the
\texttt{lambda} operator and other constructors of type \texttt{expr}.
As mentioned, using Hybrid provides the flexibility to mix both kinds
of representations.

\subsection{A Linear Specification Logic}
\label{sec:sl}
In this section, we will give a brief account of linear logic,
highlighting differences with minimal intuitionistic logic, which is
the main SL used in Hybrid historically. Then we present the sequent
calculus of our selected version of linear logic, namely an
intuitionistic linear logic.
It has both an intuitionistic and a linear context
of assumptions.  The latter is important for modeling the type system
of Proto-Quipper.

In minimal intuitionistic logic, there are three logical connectives
($\wedge$,$\vee$ and $\Rightarrow$), in addition to the logical
constants True and False.
A sequent in this logic has the form $\Gamma\vdash C$ where $\Gamma$
is a logical context (a set of formulas) and $C$ is a formula.  If a
sequent $\Gamma\vdash
C$ is valid in intuitionistic logic, then the sequent obtained by
adding a hypothesis $B$ to the context, i.e., $\Gamma,B\vdash C$, is
also valid.  This is called context weakening.
If a sequent of the form $\Gamma,B,B\vdash C$ is valid in
intuitionistic logic, then the sequent $\Gamma,B\vdash C$ is
also valid.  This is called context contraction. These two key
structural properties of intuitionistic logic are primarily prohibited
in linear logic, where we deal with the context as a collection of
resources, i.e., the hypotheses are considered as resources that can
be used one time and they must be consumed (or used). Accordingly,
$\Gamma \vdash C$ does not guarantee the linear validity of
$\Gamma,B \vdash C$, and $\Gamma,B,B \vdash C$ does not guarantee the
linear validity of $\Gamma,B \vdash C$. In addition, linear logic has
two types of logical connectives: the
multiplicative connectives ($\otimes$, $\parr$ and $\multimap$), and
the additive connectives $\&$, $\oplus$ and $\Rightarrow$. In addition
to the intuitionistic constants, there is the universal consumer
constant $\top$, which can consume any linear resource (i.e.,
hypothesis).
Our choice for SL is a slightly different version of the standard
linear logic, namely intuitionistic multiplicative linear logic, with
the two kinds of contexts mentioned above: the intuitionistic context
$\Gamma$ (a set of formulas) and the linear context $\Delta$ (a
multiset of formulas).  The contraction and weakening rules are
permitted for the intuitionistic context $\Gamma$.  The logic has two
classes of formulas---goals and program clauses---whose syntax is:
$$\begin{array}{lccl}
\mbox{Goals} & G & {}::={} & A \mid
                       \top \mid
                       \forall x:\tau\; G \mid
                       A \Rightarrow G \mid
                       G_1 \& G_2
                       \mid G_1 \otimes G_2 \mid
                       A \multimap G \\
\mbox{Clauses} & P & {}::={} & \forall
(A\longleftarrow [G_1,\ldots,G_m] [G'_1,\ldots,G'_n])
\end{array}$$
In the above grammar, $A$ is an atomic formula.  Note that this logic
includes universal quantification over typed variables.  Here, $\tau$
is a primitive type.  It will be instantiated with \texttt{qexp} in
our case study.  A clause in the form above consists of two
\emph{lists}, the first one of intuitionistic goals, the other of
linear ones. It is an abbreviation for the formula:
$$\begin{array}{l}
\forall (G_1\Rightarrow\cdots\Rightarrow G_m
\Rightarrow G'_1 \multimap\cdots\multimap G'_n \multimap A)
\end{array}$$
where the outer $\forall$ represents quantification over all free
variables in $G_1, \ldots, G_m$, $G'_1, \dots G'_n,A$,
whose types may be $\tau$ or $\tau\rightarrow\tau$, where $\tau$ is
primitive.  Thus, our logic is second-order in the sense that it is
possible to quantify over functions (of type \texttt{qexp -> qexp} in
our case).

Sequents of this logic have the form $\Gamma;\Delta\vdash_\Pi G$,
where $G$ is a goal formula, $\Gamma$ is an intuitionistic context of
formulas, $\Delta$ is a linear context, and $\Gamma$ and $\Delta$
contain only atomic formulas.  $\Pi$ is a set of program clauses,
which we omit when presenting the rules because it is fixed and does
not change within a proof.  This format emphasizes the view of this
calculus as a non-deterministic logic programming interpreter, which
is a feature of most SLs implemented in the two-level style.  See e.g.,
\cite{FeltyMomigliano:JAR12} and \cite{LProlog}.  The sequent rules of
such a logic are shown in Figure~\ref{fig:IL}.
\begin{figure}\label{tab:ll}
\centering
  \begin{tabular}{c}
\AxiomC{}
\AxiomC{}
\RightLabel{l\_init}
\BinaryInfC{$\Gamma;A\vdash A$}
\DisplayProof
\hspace{1cm}
\AxiomC{}
\AxiomC{}
\RightLabel{i\_init}
\BinaryInfC{$\Gamma,A;.\vdash A$}
\DisplayProof\\ \\

\AxiomC{$\Gamma;\Delta_1\vdash B$}
\AxiomC{$\Gamma;\Delta_2\vdash C$}
\RightLabel{$\otimes$-R}
\BinaryInfC{$\Gamma;\Delta_1,\Delta_2\vdash B\otimes C$}
\DisplayProof
\hspace{1cm}
\AxiomC{$\Gamma;\Delta\vdash B$}
\AxiomC{$\Gamma;\Delta\vdash C$}
\RightLabel{$\&$-R}
\BinaryInfC{$\Gamma;\Delta\vdash B \& C$}
\DisplayProof
\\ \\
\AxiomC{$\Gamma,A;\Delta\vdash B$}
\RightLabel{$\Rightarrow$-R}
\UnaryInfC{$\Gamma;\Delta\vdash A  \Rightarrow B$}
\DisplayProof
\hspace{.5cm}
\AxiomC{$\Gamma;\Delta,A\vdash B$}
\RightLabel{$\multimap$-R}
\UnaryInfC{$\Gamma;\Delta\vdash A  \multimap B$}
\DisplayProof

\\ \\
\AxiomC{}
\RightLabel{$\top$-R}
\UnaryInfC{$\Gamma;\Delta\vdash \top$}
\DisplayProof
\hspace{.5cm}
\AxiomC{$\Gamma;\Delta\vdash B[y/x]$}
\RightLabel{$\forall$-R}
\UnaryInfC{$\Gamma;\Delta\vdash \forall x. B$}
\DisplayProof
\\ \\
\AxiomC{$\Gamma,A,A;\Delta\vdash B$}
\RightLabel{Contraction}
\UnaryInfC{$\Gamma,A;\Delta\vdash B$}
\DisplayProof
\hspace{1cm}
\AxiomC{$\Gamma,A_1;\Delta\vdash B$}
\RightLabel{Weakening}
\UnaryInfC{$\Gamma,A_1,A_2;\Delta\vdash B$}
\DisplayProof
\\ \\
\AxiomC{$\begin{array}{l}
A \longleftarrow [G_1,\ldots,G_m][G'_1,\ldots,G'_n] \in [\Pi]\\
\Gamma; . \vdash G_i \quad (i=1,\ldots,m)\\
\Gamma;\Delta_i \vdash G'_i \quad (i=1,\ldots,n)
\end{array}$}
\RightLabel{$\mathit{bc}$}
\UnaryInfC{$\Gamma;\Delta_1,\ldots,\Delta_n\vdash A$}
\DisplayProof
\end{tabular}
\caption{Intuitionistic Linear Logic Sequent Rules \label{fig:IL}}
\end{figure}
There are two initialization rules.  The linear rule (l\_init) strictly
prohibits the existence of any hypothesis inside $\Delta$ except
\textit{A}, and it does not care about the contents of $\Gamma$. The
intuitionistic rule (i\_init) strictly requires an empty $\Delta$
whereas \textit{A} should be part of $\Gamma$. We can use $\&$ if its
operands can be proven linearly at the same time, i.e., all the
required linear resources are available in the contexts of both
premises ($\&$-R). Recall that a linear resource can be used only
once. On the other hand, additive conjunction requires that the
resources be split and used in the proof of only one premise
($\otimes$-R). This connective is suitable when the operands are
sharing the linear resources. The implication rules ($\Rightarrow$-R
and $\multimap$-R) vary based on which context the antecedent
\textit{A} comes from.
The $\forall$-R rule has the usual proviso that $y$ does not appear in
$\Gamma$, $\Delta$, or $B$.

In the $\mathit{bc}$ rule, $[\Pi]$ represents all possible instances
of clauses in $\Pi$ (clauses with instantiations for all variables
quantified at the outermost level).  Applying this rule in a backward
direction corresponds to \emph{backchaining} on a clause in $\Pi$,
instantiating the universal quantifiers so that the head of the clause
matches $A$.  There is one hypothesis for each \emph{subgoal}, both
linear and intuitionistic.

The first step towards the formalization of the linear specification
logic in Coq is defining an inductive type \texttt{oo} for formulas given by
the $G$ and $P$ grammars:
\begin{flushleft}
\texttt{Inductive~oo:~Set~:=}\\
\texttt{|~atom:~atm~-{>}~oo}\\
\texttt{|~T:~oo}\\
\texttt{|~Conj:~oo~-{>}~oo~-{>}~oo}\\
\texttt{|~And:~oo~-{>}~oo~-{>}~oo}\\
\texttt{|~Imp:~atm~-{>}~oo~-{>}~oo}\\
\texttt{|~lImp:~atm~-{>}~oo~-{>}~oo}\\
\texttt{|~All:~(expr~con~-{>}~oo)~-{>}~oo.}\\
\end{flushleft}
where the \texttt{atom} constructor accepts an atomic formula of type
\texttt{atm} and casts it into the formula type \texttt{oo}.
The type \texttt{atm} is defined for each OL and typically includes
the atomic relations or predicates of the OL, e.g., \texttt{typeof}
for Proto-Quipper typing (see Section~\ref{sec:encodepqsyntax}).
The constructor \texttt{T}
corresponds to the universal consumer, \texttt{Conj} corresponds to
multiplicative conjunction and \texttt{And} to additive
conjunction. The type constructors \texttt{Imp} and \texttt{lImp}
corresponds to intuitionistic and linear implication, respectively,
where in both cases, the formula on the left must be an atom.  The
\texttt{All} constructor takes a function as an argument, and thus the
bound variable in the quantified formula is encoded using lambda
abstraction in Coq.

The next step is defining the sequent rules themselves. This step is
done using an inductive predicate definition as follows:
\begin{flushleft}
\texttt{Inductive~seq:~nat~-{>}~list~atm~-{>}~list~atm~-{>}~oo~-{>}~Prop~~:=}\\
\texttt{|~s\_bc:~forall~(i:nat)~(A:atm)~(IL~LL:list~atm)~}\\
\texttt{~~(lL~iL:list~oo),~prog~A~iL~lL~-{>}~}\\
\texttt{~~splitseq~i~L~[]~iL~-{>}~splitseq~i~IL~LL~lL~-{>}~}\\
\texttt{~~seq~(i+1)~IL~LL~(atom~A)}\\
$\vdots$\\
\texttt{|~s\_all:~forall~(i:nat)~(B:expr~con~-{>}~oo)~(IL LL:list~atm),}\\
\texttt{~~(forall~x:expr~con,~proper~x~-{>}~seq~i~IL~LL~(B~x))~-{>}}\\
\texttt{~~seq~(i+1)~IL~LL~(All~B)}\\
\texttt{with}\\
\texttt{splitseq:~nat~-{>}~list~atm~-{>}~list~atm~-{>}~list~oo~-{>}~Prop~:=}\\
\texttt{|~ss\_init:~forall~(i:nat)~(IL:list atm),~splitseq~i~IL~[]~[]}\\
\texttt{|~ss\_general:~forall~(i:nat)~(IL~lL1~lL2~lL3:list~atm)}\\
\texttt{~~(G:oo)~(Gs:list oo),~}\\
\texttt{~~split~lL1~lL2~lL3~-{>}~seq~i~IL~lL2~G~-{>}~}\\
\texttt{~~splitseq~i~IL~lL3~Gs~-{>}~splitseq~i~IL~lL1~(G::Gs).}\\
\end{flushleft}
where we use the variables \texttt{IL} to represent the intuitionistic
context, and \texttt{LL} for the linear one. The definition of
\texttt{seq} follows the standard rules described earlier.  Only a
representative set of the inference rules of the formal definition is
shown above.
The natural number argument allows proofs by induction over the height
of a sequent derivation, which we often use.  This argument can be
ignored when doing proofs by structural induction.
The reader is referred to~\cite{ourformal} for the full
definition. In the \texttt{s\_bc} rule the predicate \texttt{splitseq}
is used to check the provability of a list of subgoals. The predicate
\texttt{splitseq} is used twice; once for the intuitionistic subgoals
\texttt{iL} under the empty linear context, and once for the
linear subgoals \texttt{lL}.
When we discuss the rules for \texttt{prog} later, we will say that we
must \textit{intuitionistically prove} the goals in \texttt{iL} and
\textit{linearly prove} the goals in \texttt{lL}.
The predicates \texttt{seq} and
\texttt{splitseq} are defined using Coq's mutual induction.  The
inductive definition of \texttt{splitseq} is at the very end of the
definition.  In the case when the list of goals is non-empty, the head
subgoal \texttt{G} is proven under the linear context \texttt{lL2} and
the remaining list of subgoals \texttt{Gs} is proven under the context
\texttt{lL3} if and only if there exists a context \texttt{lL1} such
that \texttt{split lL1 lL2 lL3}.  We describe the \texttt{split}
predicate by example (and omit its definition here): the split of
[A;B;C] can be [A;B] and [C], [C] and [A;B], [A] and [C;B], or [B] and
[C;A]. The idea of the split is that the two sublists divide the big
list, regardless of the order of elements inside the sublist.  This is
to ensure that there are no shared linear resources between
\texttt{lL2} and \texttt{lL3}. Finally, \texttt{(prog A iL lL)}
represents a formula from the program context defining the rules of
the OL (i.e., $\Pi$).  Later (in
Section~\ref{sec:encodepqsyntax}), we will see the implementation of
\texttt{prog} for Proto-Quipper.

The rule for \texttt{All} has an argument of function type
$\mathtt{expr}~\mathtt{con}\to \mathtt{oo}$, unlike the other
rules,
since quantification is over terms of the OL.  Recall
that \texttt{con} is a parameter to the type \texttt{expr}, and must
be implemented for each OL.  Note that we restrict
\texttt{x} to terms satisfying the \texttt{proper} predicate (terms
without dangling indices).

The implemented specification logic has been validated by proving a
number of essential properties. We show two here. The first property
is a cut-elimination rule for the intuitionistic context:
\begin{flushleft}
\texttt{Lemma~seq\_cut\_aux:}\\
\texttt{~~forall~(i~j:nat)~(a:atm)~(b:oo)~(il~ll:list~atm),}\\
\texttt{~~seq~i~il~ll~b~-{>}~In~a~il~-{>}}\\
\texttt{~~seq~j~(remove~eq\_dec~a~il)~[]~(atom~a)~-{>}~}\\
\texttt{~~seq~(i+j)~(remove~eq\_dec~a~il)~ll~b).}\\
\end{flushleft}
\noindent The theorem states that if we remove all instances of the
hypothesis \texttt{a} from the list of intuitionistic hypotheses
\texttt{il}, and \texttt{a} is found to be provable under the new list
of hypotheses, then eliminating \texttt{a} does not affect the
provability of \texttt{b}. Note that the \texttt{remove} function
takes a proof that equality at type \texttt{atm} is decidable.  Since
\texttt{atm} is a parameter to the SL, so is \texttt{eq\_dec}. The
second property is the weakening property for the intuitionistic
context:
\begin{flushleft}
\texttt{Theorem~seq\_weakening\_cor:}\\
\texttt{~~forall~(i:nat)~(b:oo)~(il~il'~ll:list~atm),}\\
\texttt{~~(forall~(a:atm),~In~a~il~-{>}~In~a~il')~-{>}}\\
\texttt{~~seq~i~il~ll~b~-{>}~seq~i~il'~ll~b.}\\
\end{flushleft}
This concludes the formalization of the linear specification language
and some of its metatheory. In the following sections, we will
present Proto-Quipper as an OL that benefits from this
logic, where will give a concrete implementation for the parameters
presented in this section, in particular, \texttt{atm}, \texttt{con},
and \texttt{prog}, as well as prove decidability of equality for our
instantiation of \texttt{atm}.

\section{Encoding Proto-Quipper Programs and Semantics in Hybrid}
\label{sec:encoding}
In this section, we discuss
a key aspect
of this work,
where we present the encoding of Proto-Quipper in the Hybrid
framework. This includes the formalization of Proto-Quipper syntax as
a concrete implementation of the types \texttt{con} and \texttt{atm},
and the main parts of the program context \texttt{prog}, which includes the typing
rules in Figure \ref{fig:typingrules}. We then present a number of theorems  that are important for proving type soundness of Proto-Quipper.

\subsection{Encoding Proto-Quipper Terms}
\label{sec:pqsyntax}
Recall that in Hybrid, \texttt{con} is a parameter to the type
\texttt{expr}. The implementation of \texttt{con} for an OL
typically includes all constants that appear in the
language. This includes the names of the operations supported by the
OL, e.g., \texttt{if} and \texttt{let}. The
implementation of \texttt{con} for Proto-Quipper is as follows:
\begin{flushleft}
\texttt{Inductive~Econ:~Set~:=}\\
\texttt{|~qABS:~Econ~|~qAPP:~Econ~|~qPROD:~Econ~}\\
\texttt{|~qLET:~Econ~|~sLET:~Econ~|~qCIRC:~Econ~|~qIF:~Econ}\\
\texttt{|~BOX:~qtp -> Econ~|~UNBOX:~Econ~|~REV:~Econ}\\
\texttt{|~TRUE:~Econ~|~FALSE:~Econ~|~STAR:~Econ}\\
\texttt{|~Qvar:~nat~-{>}~Econ~|~Crcons:~nat~-{>}~Econ.}\\
\end{flushleft}
This definition might cause some confusion since it does not
differentiate between constants like \texttt{TRUE} and \texttt{FALSE},
and operations like lambda abstraction \texttt{qABS} and the let
statement \texttt{qLET}. Actually, these are just constants and this
definition does not include any semantics or functionality by
itself. The functionality of these operations will be addressed
next. In our formalization, we encode all quantum variables of
Proto-Quipper as constants using the constant \texttt{Qvar}, which
maps natural numbers to quantum variables. Similarly, the constant
\texttt{Crcons} models the names of quantum circuits, i.e., the
circuit constant $C$ (see Section \ref{sec:pq}). The argument of type
\texttt{qtp} to the \texttt{BOX} operator encodes the type superscript
in $\mathit{box}^T$.  As mentioned earlier, \texttt{qexp} is an
abbreviation for \texttt{(expr Econ)}.
\begin{flushleft}
\texttt{Definition~qexp:~Set~:=~expr~Econ.}
\end{flushleft}

As stated earlier, all OL constructs are encoded using
\texttt{CON}, \texttt{VAR}, \texttt{APP} and \texttt{lambda}. These
definitions are the only place where the constructors of the
\texttt{expr} type and the \texttt{lambda} operator are seen explicitly.
Once the new constants are defined, we use only these.  We start with
the simplest item in the language, namely variables:
\begin{flushleft}
\texttt{Definition~Var:~var~-{>}~qexp~:=~VAR Econ}
\end{flushleft}
\noindent The definition is simple as it just uses the Hybrid
\texttt{VAR} parameterized with the Proto-Quipper constants
\texttt{Econ}.
Another
simple example of encoding Proto-Quipper operations in Hybrid is function
application:
\begin{flushleft}
\texttt{Definition~App~(e1~e2:qexp)~:~qexp~:=}\\
\texttt{~~APP~(APP~(CON~qAPP)~e1)~e2.}
\end{flushleft}
\noindent To help the reader to understand the above definition, it is
better to view the Hybrid constructor \texttt{APP} as a concatenation
operator. Another possible definition is
\texttt{(APP (CON qAPP) (APP e1 e2))}. One might wonder if it is useless
to add the constant \texttt{qAPP}. Remember that the \texttt{APP}
constructor is used to represent other program statements, e.g., the \texttt{if}
statement. Therefore, we need to add such constants for each type of expression
so we can identify statements with different meanings. Similarly, we
formally define the product and circuit statements as follows:
\begin{flushleft}
\texttt{Definition~Prod~(e1~e2:qexp)~:~qexp~:=}\\
\texttt{~~APP (APP (CON qPROD) e1) e2.}\\
\medskip
\texttt{Definition~Circ~(e1:qexp)~(i:nat)~(e2:qexp)~:~qexp~:=}\\
\texttt{~~APP~(APP~(APP~(CON~qCIRC)~e1)~(CON~(Crcons~i)))~e2}.
\end{flushleft}
Now, let us  turn to a Proto-Quipper construct that is a bit more difficult,
namely function abstraction.
\begin{flushleft}
\texttt{Definition~Fun~(f:qexp~-{>}~qexp)~:~qexp~:=}\\
\texttt{~~APP~(CON~qABS)~(lambda~f).}\\
\end{flushleft}
\noindent This expression has an argument that is a function of type
\texttt{(qexp~-{>}~qexp)}. To better understand this construct, we
give some examples.  Let \texttt{f} be the function \texttt{fun
  x~={>}~App (Var 0) x}. Consider the term \texttt{(lambda f)}.
If definitions are expanded, the Hybrid operator \texttt{lambda}
disappears and the result is the de Bruijn format of this term which
replaces the bound variable \texttt{x} by the correct de Bruijn
index. For the above example, \texttt{(lambda f)} unfolds to
\texttt{ABS~(APP (VAR ECON 0) (BND 0))}. As stated, we never need to
expand such definitions.  Now that we have the full encoding of
Proto-Quipper terms, we can return to the example term from the
introduction, $\lambda x . \lambda y . x y$, which is
encoded as \texttt{(Fun~(fun~x:qexp~=>~(Fun~(fun~y:qexp~=>~(App~x~y)))))}.

Similarly, we handle the more
difficult case of the \texttt{Let} expression.
\begin{flushleft}
\texttt{Definition~Let~(f:qexp~->~qexp~->~qexp)~(e1:qexp)~:~qexp~:=}\\
\texttt{~~APP~(CON~qLET)~(APP~(lambda~(fun~x~=>~(lambda~(f~x))))~e1)}.
\end{flushleft}
Recall that the \texttt{let} statement in Proto-Quipper is
restricted to product expressions, i.e., it always takes the form
$\mathtt{let}~\langle x, y \rangle = a~\mathtt{in}~b$.  Therefore, we have two
function abstractions, e.g., \texttt{fun~x~=>~fun~y~=>~App y x}. Since the
\texttt{lambda} is only defined for functions of type
\texttt{exp~-{>}~exp} (not \texttt{exp~-{>}~exp~-{>}~exp}), we have to
make two applications of \texttt{lambda} in the way presented in the above
definition in order to satisfy the typing condition of
\texttt{lambda}.
This is the first case study using Hybrid that requires a function of
more than one argument to represent OL syntax.

We note
that terms in Hybrid are equivalent up to $\eta$-conversion, so the
body of the definition of \texttt{Fun} and \texttt{Let} operators,
respectively, could also be written:
\begin{flushleft}
\texttt{(APP~(CON~qABS)~(lambda~(fun~x~={>}~(f~x)))).}\\
\texttt{(APP~(CON~qLET)}\\
\texttt{~~(APP~(lambda~(fun~x~={>}~(lambda~(fun~y~=>~(f~x~y)))))~b)).}
\end{flushleft}

The formal encoding of the other Proto-Quipper expressions is  similar to the definitions presented above \cite{ourformal}.

We also define the following predicate that holds for expressions
that only involve quantum variables, the star constant, and the
product constructor \texttt{Prod}.
\begin{flushleft}
\texttt{Inductive~quantum\_data:~qexp~-{>}~Prop~:=~}\\
\texttt{~~vQVAR:~forall~i,~quantum\_data~(CON~(Qvar~i))}\\
\texttt{|~vSTAR:~quantum\_data~(CON~STAR)}\\
\texttt{|~vTENSOr:~forall~a~b,~quantum\_data~a~-{>}~quantum\_data~b~-{>}}\\
\texttt{~~quantum\_data~(Prod~a~b).~}\\
\end{flushleft}
This subset of expressions are those whose types satisfy the
\texttt{valid} predicate.

\subsection{Encoding Proto-Quipper Semantics}
\label{sec:encodepqsyntax}
The main purpose of this section is to discuss the implementation of
the atomic predicates \texttt{atm} and the program context \texttt{prog}
for Proto-Quipper.  Here, \texttt{atm} contains three constructors,
one that relates a Proto-Quipper type \texttt{qtp} to a Proto-Quipper
expression \texttt{qexp}, one for representing reduction rules, and
one that identifies valid (well-formed) expressions of Proto-Quipper.
\begin{flushleft}
\texttt{Inductive~atm~:~Set~:=}\\
\texttt{|~typeof~:~qexp~-{>}~qtp~-{>}~atm}\\
\texttt{|~reduct~:~Econ~->~qexp~-{>}~Econ~->~qexp~-{>}~atm}\\
\texttt{|~is\_qexp~:~qexp~-{>}~atm.}\\
\end{flushleft}
It is important to clarify that the formalization presented
in the previous section does not guarantee  that all instances of
the type \texttt{qexp} are valid in Proto-Quipper. (It just defines
expressions that we are interested in.) This will be done as part of
the program context \texttt{prog} with the help of the constructor
\texttt{is\_qexp}.

Now, we turn to the crucial step in our whole formalization, the
implementation of \texttt{prog}. The formal definition of such a
predicate is quite long. Accordingly, we choose to present it rule by
rule, where we address only the major rules, and others can be found
in~\cite{ourformal}.  We discuss clauses for \texttt{is\_qexp} and
\texttt{typeof} here, and leave the discussion of \texttt{reduct} to
Section~\ref{sec:sr}.

Now that we have instantiated the
parameters \texttt{con} and \texttt{atm}, we can define the first two
abbreviations below.
\begin{flushleft}
\texttt{Definition~oo\_~:=~oo~atm~Econ.}\\
\texttt{Definition~atom\_:~atm~->~oo\_~:=~atom~Econ.}\\
\texttt{Definition~seq\_:~nat~->~list~atm~->~list~atm~->~oo\_~->~Prop~:=~}\\
\texttt{~~seq prog.}\\
\texttt{Definition~splitseq\_:}\\
\texttt{~~nat~->~list~atm~->~list~atm~->~list~oo\_~->~Prop~:=}\\
\texttt{~~splitseq prog.}
\end{flushleft}
The third and fourth are useful once we have completed the definition of
\texttt{prog}.  Thus the general form of sequents of the SL will be
written \texttt{seq\_ n IL LL G}, where the arguments are the height
of the proof, the intuitionistic and linear contexts of assumptions of
type \texttt{atm}, and the conclusion (formula on the right of the
turnstile), respectively.  We will sometimes omit the height argument
for readability when it is not important for the discussion, and in
particular in the statements of theorems, \texttt{seq\_ IL LL G} will
mean that there exists an \texttt{n} such that \texttt{seq\_ n IL LL
  G}.

The \texttt{prog} predicate has type
\texttt{~atm~-{>}~list~oo\_~-{>}~list~oo\_} \texttt{~-{>}~Prop}, whose
arguments are: the atomic statement, the list of intuitionistic
subgoals, and the list of linear subgoals.  Recall that it appears in
the definition of \texttt{seq} in the \texttt{s\_bc} clause, which
implements the $\mathit{bc}$ rule in Figure \ref{fig:IL}.  It is
the implementation of the program clauses $\Pi$.  In the
upcoming rules, this predicate reads as follows: the atomic statement
is valid in the context of Proto-Quipper program if the list
of intuitionistic and linear subgoals can be proven using the linear
specification logic presented in Figure \ref{fig:IL} (as implemented
by \texttt{seq}). We start by presenting examples of syntax rules that
define valid expressions inside Proto-Quipper:
\begin{flushleft}
\texttt{|~starq:~prog~(is\_qexp~(CON~STAR))~[]~[]}\\
\texttt{|~trueq:~prog~(is\_qexp~(CON~TRUE))~[]~[]}\\
\texttt{|~boxq:~prog~(is\_qexp~(CON~BOX))~[]~[]}\\
\end{flushleft}
Constants of Proto-Quipper are unconditionally valid; the list of
linear and intuitionistic subgoals are empty. Similar clauses are
included in \texttt{prog} for the constants \texttt{FALSE},
\texttt{UNBOX}, and \texttt{REV}.

Note that we have defined the syntax of terms in Section~\ref{sec:pq}
as grammars.  As argued in~\cite{FMP:MSCS17}, grammars contain implicit
information, and specifying well-formedness as inference rules makes
some of this information explicit, which is useful for formally
defining valid terms. For example, consider the
sublanguage of Proto-Quipper terms containing term variables,
application, and function abstraction.  One possible set of rules
expressing valid syntax is as follows:
$$\begin{array}{c}
\infer{\Phi \vdash \mathit{is\_qexp}~{x}}{\mathit{is\_qexp}~{x} \in \Phi}
\qquad
\infer{\Phi \vdash \mathit{is\_qexp}~ab}
      {\Phi \vdash \mathit{is\_qexp}~a\qquad  \Phi \vdash \mathit{is\_qexp}~b}\\[1em]
\infer{\Phi \vdash \mathit{is\_qexp}~\lambda x.a}
      {\Phi, \mathit{is\_qexp}~x \vdash \mathit{is\_qexp}~a}
  \end{array}$$
Here, $\Phi$ is an explicit context of variables, and a term is only
considered well-formed if all of its free variables come from this
set.  Stated formally in our setting, for a sequent \texttt{seq\_ IL
  [] (atom\_ (is\_qexp a))} to be provable, for each free variable
\texttt{x} of type \texttt{qexp} that appears in \texttt{a}, the atom
\texttt{is\_qexp x} must be in the intuitionistic context \texttt{IL}.
Note that the list of linear subgoals is empty. This is the
case for all well-formedness of syntax rules, as the context elements
expressing validity of an expression can be used as many times as we
want, i.e., they are infinitely consumable resources.

The following clause encodes the well-formedness of the application
expression:
\begin{flushleft}
\texttt{|~apq:~forall~E1~E2:qexp,}\\
\texttt{~~prog~(is\_qexp~(App~E1~E2))}\\
\texttt{~~~[And~(atom\_~(is\_qexp~E1))~(atom\_~(is\_qexp~E2))]~[]}\\
\end{flushleft}
As mentioned, the list of linear subgoals is empty.  In the
intuitionistic goals,
notice is the use of additive conjunction. Actually, we cannot use
multiplicative conjunction to express an intuitionistic subgoal
according to the specification logic defined in Section
\ref{sec:sl}. The use of \texttt{And} for intuitionistic subgoals has
the same power as the use of multiplicative conjunction because the
use of the intuitionistic context involves infinitely consumable
resources, and hence we can show the validity of \texttt{E1} and
\texttt{E2} from the same context. Similar rules are defined for the
\texttt{Prod} and \texttt{Slet} expressions, and also the \texttt{If}
expression, where we have three sub-expressions instead of two.

The following clause
corresponds to the function abstraction (\texttt{Fun}) case.
\begin{flushleft}
\texttt{|~lambdaq:~forall~(E:qexp~-{>}~qexp),~abstr~E~-{>}}\\
\texttt{~~prog~(is\_qexp~(Fun~E))}\\
\texttt{~~~[All~(fun~x:qexp~={>}~}\\
\texttt{~~~~~~~~~~Imp~(is\_qexp~x)~(atom\_~(is\_qexp~(E~x))))]~[]}
\end{flushleft}
For a function expression containing \texttt{E} to be valid in Proto-Quipper, \texttt{E} should
first satisfy the Hybrid \texttt{abstr} condition. This predicate
guarantees that \texttt{E} encodes object-level syntax (discussed
earlier and in more detail in Section~\ref{sec:adequacy}). One of the
big advantages of the Hybrid framework is that it hides such details
when reasoning about OLs. The term \texttt{(Fun E)} is said
to be valid if for all valid expressions \texttt{x}, the
expression \texttt{E x} a is valid Proto-Quipper expression.
Again, this subgoal needs to be proved intuitionistally.
Proving this subgoal involves assuming that a new variable \texttt{x}
is a valid expression and showing that the term obtained by replacing
the bound variable in \texttt{E} by \texttt{x} is a valid expression.
A similar clause can be developed for the \texttt{Let}
expression, with the difference that \texttt{E} is a function of two
parameters instead of one:
\begin{flushleft}
\texttt{|~letq:~forall~(E:qexp~-{>}~qexp~-{>}~qexp)~(b:qexp),}\\
\texttt{~~abstr~(fun~x~={>}~lambda~(E~x))~-{>}}\\
\texttt{~~abstr~(fun~y~={>}~lambda~(fun~x~={>}~(E~x~y)))~-{>}}\\
\texttt{~~prog~(is\_qexp~(Let~E~b))}\\
\texttt{~~~[All~(fun~x~:~qexp~={>}~(All~(fun~y:qexp~={>}~}\\
\texttt{~~~~~~~~~~Imp~(is\_qexp~x)~(Imp~(is\_qexp~y)~}\\
\texttt{~~~~~~~~~~~~~~~~(atom\_~(is\_qexp~(E~x~y)))))));}\\
\texttt{~~~~atom\_~(is\_qexp~b)]}\\
\texttt{~~~[]}\\
\end{flushleft}
This case illustrates how \texttt{abstr} is used for functions of more
than one argument.  Another difference from the \texttt{Fun} case is
that a second subgoal ensures the validity of the \texttt{b} argument
representing the body of the $\mathtt{let}$ expression.

The last syntactic rule to present is the quantum circuit rule
\texttt{circ}.  Note that in the sample rules for well-formedness of
Proto-Quipper expressions given above, the context $\Phi$ should
contain all the free term variables; we did not include a second
context for quantum variables.  Following the style of the typing rule
$\mathit{circ}$, we use the $In$ and $Out$ functions to identify free
quantum variables and then these variables occur in the corresponding
premises, but not in the conclusion.  In the rule below, if $Q$ is a
set of quantum variables $q_1,\ldots q_n$ for $n\ge0$, we write
$\Phi_Q$ to abbreviate the context
$\mathit{is\_qexp}~q_1,\ldots,\mathit{is\_qexp}~q_n$.
$$\begin{array}{c}
\AxiomC{$\Phi,\Phi_{Q_1} \vdash \mathit{is\_qexp}~t$
  \hspace{.5cm}
  $\Phi, \Phi_{Q_2} \vdash \mathit{is\_qexp}~a$
  \hspace{.5cm}
  $In(C) = Q_1$
  \hspace{.5cm}
  $Out(C)=Q_2$}
\RightLabel{}
\UnaryInfC{$\Phi \vdash \mathit{is\_qexp}~(t,C,a)$}
\DisplayProof
\end{array}$$
This rule is encoded as the following clause of the \texttt{prog}
definition.
\begin{flushleft}
\texttt{|~Circq:~forall~(C:nat)~(t~a:qexp),~quantum\_data~t~-{>}}\\
\texttt{~~prog~(is\_qexp~(Circ~t~C~a))}\\
\texttt{~~~[toimpexp~(FQ~a)~(atom\_~(is\_qexp~a))]~[]}\\
\end{flushleft}
Recall that \texttt{quantum\_data} is a predicate that holds for
expressions that only involve quantum variables, the star constant,
and the product constructor \texttt{Prod}.  This predicate is used to
directly prove the first premise $\Phi,\Phi_{Q_1} \vdash
\mathit{is\_qexp}~t$. The encoding of the second premise appears in
the intuitionistic list of subgoals (the second argument to
\texttt{prog}). The Coq function \texttt{FQ} returns a list of free
quantum variables of an expression, and \texttt{toimpexp} is a
function that takes a list quantum variables and using intuitionistic
implication, adds them recursively as antecedents to a the predicate
expressing well-formedness of the term. We illustrate by example:
\begin{flushleft}
\texttt{toimpexp~(FQ~(Prod~(CON~Qvar~0)~(CON~Qvar~1)))}\\
\texttt{~~~~~~~~~(atom\_~(is\_qexp~(Prod~(CON~Qvar~0)~(CON~Qvar~1))))~=}\\
\texttt{Imp~(is\_qexp~(CON~Qvar~0)}\\
\texttt{~~(Imp~(is\_qexp~(CON~Qvar~1))}\\
\texttt{~~~~~(atom\_~(is\_qexp~(Prod~(CON~Qvar~0)~(CON~Qvar~1))))))}
\end{flushleft}
In general, proving such an implication requires repeated applications of the
$\Rightarrow$-R rule of the SL (Figure~\ref{fig:IL}) in a backward
direction, moving antecedents of the form $\mathit{is\_qexp}~q$ into
the intuitionistic context, resulting in a context that encodes
$\Phi_{Q_2}$.

As stated above, we have defined \texttt{FQ} as a Coq function.  In
general, users of Hybrid need to take care when defining functions
with arguments whose types instantiate \texttt{expr} (defined in
Section~\ref{sec:hybrid}), which in our specific case includes the
type \texttt{qexp} (defined as \texttt{expr Econ} in
Section~\ref{sec:pqsyntax}).  In particular, proving properties of
such functions may require exposing the de Bruijn representation of
these arguments.  The \texttt{FQ} function is one such example.  In
particular, our proofs require the following three axioms:
\begin{flushleft}
\texttt{Hypothesis FQ\_FUN: forall i E,}\\
\texttt{~~abstr E -> FQ (Fun E) = FQ (E (Var i)).}\\
\texttt{Hypothesis FQ\_LET: forall i E b,}\\
\texttt{~~abstr (fun x => lambda (E x)) ->}\\
\texttt{~~(forall x, proper x -> abstr (E x)) ->}\\
\texttt{~~FQ (Let E b) = (FQ (E (Var i) (Var i))) ++ (FQ b).}\\
\texttt{Hypothesis FQU\_LET: forall i E b,}\\
\texttt{~~abstr (fun x => lambda (E x)) ->}\\
\texttt{~~(forall x, proper x -> abstr (E x)) ->}\\
\texttt{~~FQU (Let E b) = (FQU (E (Var i) (Var i))) ++ (FQU b).}
\end{flushleft}
These axioms are admissible, but
require complex proofs.  Replacing the definition of the \texttt{FQ}
function with an inductive predicate that relates the input and output
would solve this problem.\footnote{This is straightforward change, but
  it affects a large portion of the proof development and is left for
  (very near) future work.}

This concludes the presentation of the rules for well-formedness of
Proto-Quipper terms.  Such clauses actually play two roles in proofs
of OL meta-theory.  First, induction on well-formedness derivations is
often useful, since induction directly on terms is not currently
available in Hybrid.  Second showing that rules are adequately
represented requires these clauses.  (See Section~\ref{sec:adequacy}.)

In the following, we will present the formal typing rules that
correspond to the sequent rules presented in
Figure~\ref{fig:typingrules}.  Recall that the Proto-Quipper typing
judgment has the form $\Phi,\mathop{!}\Psi;Q\vdash a:A$.  The
latter context $Q$ contains all free quantum variables that appear on
the right of a sequent; they appear in such a context without
specifying the type since it is implicitly known to be
\texttt{qubit}. The other context contains term variables and their
types and has two parts where the types in $\mathop{!}\Psi$ contain
a leading $\mathop{!}$ and the types in $\Phi$ do not.  Recall also
that the sequents of the SL, which we will now use to encode these
rules, have the general form $\Gamma;\Delta\vdash A$, where $A$ is a
formula of the SL, $\Gamma$ is an intuitionistic context of atomic
formulas, and $\Delta$ is a linear context, also of atomic formulas.
In our formalization, elements of the contexts of the Proto-Quipper
typing judgment will be encoded as atomic formulas of the form
\texttt{(typeof x A)} in the SL; in particular, the encoding of
elements of $\mathop{!}\Psi$ will appear in the intuitionistic context
$\Gamma$ of the SL, while elements of the typing context
$\Phi$ will appear in the linear context $\Delta$ of the
SL.  Each quantum variable $q$ in $Q$ will be placed in $\Delta$
explicitly associated with the type \texttt{qubit}, i.e.,
\texttt{typeof (CON (Qvar qi)) qubit}, where \texttt{qi} is the
natural number encoding variable $q$.

We start with the $ax_c$ rule, which we specify as two clauses
depending on whether or not the first argument to the subtype relation
has a leading \texttt{bang}.
\begin{flushleft}
\texttt{|~axc1:~forall~(A~B:qtp)~(x:qexp),~validT~(bang~A)~-{>}}\\
\texttt{~~Subtyping~A~B~-{>}}\\
\texttt{~~prog~(typeof~x~B)~[atom\_~(is\_qexp~x)]~[atom\_~(typeof~x~A)]}\\
\texttt{|~axc2:~forall~(A~B:qtp)~x,~Subtyping~(bang~A)~B~-{>}}\\
\texttt{~~prog~(typeof~x~B)}\\
\texttt{~~~[(And~(atom\_~(typeof~x~(bang~A)))~(atom\_~(is\_qexp~x)))]~[]}\\
\end{flushleft}
In \texttt{axc1}, the \texttt{validT (bang A)} condition ensures that
\texttt{A} has no leading \texttt{bang}, and as a consequence of
theorem \texttt{Subtyping\_Prop6}, \texttt{B} also has no leading
\texttt{bang}.  Thus it is required to linearly prove that
\texttt{atom\_~(typeof~x~A)}, whereas in \texttt{axc2}, it is required
to intuitionistically prove \texttt{atom\_~(typeof~x~(bang A))}.  Note
that for reasons of adequacy, there is the additional proof obligation
to show that \texttt{x} is a valid Proto-Quipper expression (in both
cases).

Regarding the $ax_q$ and $ax_x$ rules, as mentioned they are not
encoded as clauses of \texttt{prog}.  In particular, they are already
covered by the initial rule of the SL (l\_init of Figure~\ref{fig:IL}
encoded as part of the defintion of \texttt{seq}).

The following clauses implement the two versions of the $\top$ rule,
the first for $n=0$ and the second for $n=1$, and the $\mathit{cst}$
rule for $\mathit{box^T}$.  Similar to the well-formedness rules for
constants, they are axioms and thus have no subgoals.
\begin{flushleft}
\texttt{|~truel:~prog~(typeof~(CON~TRUE)~bool)~[]~[]}\\
\texttt{|~truei:~prog~(typeof~(CON~TRUE)~(bang~bool))~[]~[]}\\
\texttt{|~box:~forall~T~U~B,~valid~T~-{>}~valid~U~-{>}~}\\
\texttt{~~Subtyping~(bang~(arrow~(bang~(arrow~T~U))~}\\
\texttt{~~~~~~~~~~~~~~~~~~~~~~~~~(bang~(circ~T~U))))~B~-{>}~}\\
\texttt{~~prog~(typeof~(CON~BOX)~B)~[]~[]}\\
\end{flushleft}
The \texttt{prog} definition includes similar clauses for the other
rules about constants and functions on circuits, i.e, the two versions
of the $\ast_{i}$ and $\bot$ rules, and the $\mathit{cst}$ rules for
$\mathit{unbox}$ and $\mathit{rev}$.

The next two clauses encode the $\lambda_1$ rules. Similar to the
$ax_c$ rule, we develop intuitionistic and linear versions of
$\lambda_1$, depending on the type of the bound variable (whether or
not it has a leading $\mathop{!}$):
\begin{flushleft}
\texttt{|~lambda1l:~forall~(T1~T2:qtp)~(E:qexp~-{>}~qexp),}\\
\texttt{~~abstr~E~-{>}~validT~(bang~T1)~-{>}~validT~T2~-{>}~}\\
\texttt{~~prog~(typeof~(Fun~(fun~x~={>}~E~x))~(arrow~T1~T2))~[]~~}\\
\texttt{~~~[(All~(fun~x:qexp~={>}~Imp~(is\_qexp~x)~}\\
\texttt{~~~~~(lImp~(typeof~x~T1)~(atom\_~(typeof~(E~x)~T2)))))]}\\
\texttt{|~lambda1i:~forall~(T1~T2:qtp)~(E:qexp~-{>}~qexp),}\\
\texttt{~~abstr~E~-{>}~validT~(bang~T1)~-{>}~validT~T2~-{>}}\\
\texttt{~~prog~(typeof~(Fun~(fun~x~={>}~E~x))~(arrow~(bang~T1)~T2))~[]}\\
\texttt{~~~[(All~(fun~x:qexp~={>}~Imp~(is\_qexp~x)~}\\
\texttt{~~~~~(Imp~(typeof~x~(bang~T1))~(atom\_~(typeof~(E~x)~T2)))))]}\\
\end{flushleft}
Note that in both cases, the subgoal is required to be proved linearly
regardless of the type of the bound variable. This is because the type
of the whole expression \texttt{Fun (fun x => E x)} is linear. The
difference between the two rules is the use of \texttt{lImp} when the
type of the bound variable is linear, and the use of \texttt{Imp} when
the type of the bound variable is duplicable. In contrast, the subgoal
will be required to be proved intuitionistically for both cases of the
$\lambda_2$ rule. We omit the clauses for these rules, as well as
those for the $\otimes_e$ rule,
since they are similar to the encoding of $\lambda_1$ presented above.

The following clause implements the $\mathit{app}$ rule. This case
requires that both expressions be available at the same
time. Accordingly, it uses multiplicative conjunction (\texttt{Conj}):
\begin{flushleft}
\texttt{|~tap:~forall~E1~E2:qexp,~forall~T~T':qtp,~}\\
\texttt{~~validT~(arrow~T~T')~-{>}~prog~(typeof~(App~E1~E2)~T)~[]}\\
\texttt{~~[(Conj~(atom\_~(typeof~E1~(arrow~T'~T)))~}\\
\texttt{~~~~~~~~~(atom\_~(typeof~E2~T')))]}\\
\end{flushleft}
In other words, considering backward proof from the $\mathit{app}$
rule of Figure~\ref{fig:typingrules}, the linear context of the
conclusion must be divided into two disjoint contexts.  The use of
\texttt{Conj} means that the $\otimes$-R rule of Figure~\ref{fig:IL}
will be used to achieve this division.  Note here that there is only
one $\mathit{app}$ rule, and thus only one corresponding clause in
\texttt{prog} since, whether or not \texttt{T} has a leading
\texttt{bang}, the typing judgments for \texttt{E1} and \texttt{E2}
must be proved linearly.

The encoding of the two $\otimes_i$ rules is similar to
$\mathit{app}$:
\begin{flushleft}
\texttt{|~ttensorl:~forall~E1~E2:qexp,~forall~T~T':qtp,}\\
\texttt{~~validT~(tensor~T~T')~-{>}~}\\
\texttt{~~prog~(typeof~(Prod~E1~E2)~(tensor~T~T'))~[]}\\
\texttt{~~~[Conj~(atom\_~(typeof~E1~T))~(atom\_~(typeof~E2~T'))]}\\
\texttt{|~ttensori:~forall~E1~E2:qexp,~forall~T~T':qtp,}\\
\texttt{~~validT~(bang~T)~-{>}~validT~(bang~T')~-{>}}\\
\texttt{~~prog~(typeof~(Prod~E1~E2)~(bang~(tensor~T~T')))~[]}\\
\texttt{~~~[Conj~(atom\_~(typeof~E1~(bang~T)))~}\\
\texttt{~~~~~~~~~(atom\_~(typeof~E2~(bang~T')))]}\\
\end{flushleft}
We omit the clauses for $\ast_e$ and $\mathit{if}$, since their
encoding is similar to rules already presented.

The last rule is $\mathit{circ}$.
We encode this rule similarly to the encoding of the well-formedness
rule for circuits: in this case
the lists of free variables $Q_1$ and $Q_2$ have been moved in front
of the turnstile symbol with the help of linear implication for
\texttt{typeof} atoms, and intuitionistic implication for the
\texttt{is\_qexp} atoms:
\begin{flushleft}
\texttt{|~tCricl:~forall~(C:nat)~(t~a:qexp),~forall~T~U,}\\
\texttt{~~circIn~(Crcons~C)~=~FQ~t~-{>}}\\
\texttt{~~circOut~(Crcons~C)~=~FQ~a~-{>}}\\
\texttt{~~quantum\_data~t~-{>}~validT~(circ~T~U)~-{>}}\\
\texttt{~~~prog~(typeof~(Circ~t~C~a)~(circ~T~U))~}\\
\texttt{~~~~[And~(toimp~(FQ~a)~(atom\_~(typeof~a~U)))~}\\
\texttt{~~~~~~(toimp~(FQ~t)~(atom\_~(typeof~t~T)))]~[]}\\
\texttt{|~tCrici:~forall~(C:nat)~(t~a:qexp),~forall~T~U,}\\
\texttt{~~circIn~(Crcons~C)~=~FQ~t~-{>}}\\
\texttt{~~circOut~(Crcons~C)~=~FQ~a~-{>}}\\
\texttt{~~quantum\_data~t~-{>}~validT~(circ~T~U)~-{>}}\\
\texttt{~~~prog~(typeof~(Circ~t~C~a)~(bang~(circ~T~U)))}\\
\texttt{~~~~[And~(toimp~(FQ~a)~(atom\_~(typeof~a~U)))~}\\
\texttt{~~~~~~(toimp~(FQ~t)~(atom\_~(typeof~t~T)))]~[]}\\
\end{flushleft}
\noindent Similar to the \texttt{toimpexp} presented earlier, we
define \texttt{toimp} which iteratively adds well-formedness and
typing assumptions for free quantum variables to
a statement with the help of linear implication.
Again, we illustrate by example:
\begin{flushleft}
\texttt{toimp~(FQ~(Prod (CON (Qvar 0)) (CON (Qvar 1))))}\\
\texttt{~~~~~~(atom\_~(typeof~(Prod (CON (Qvar 0)) (CON (Qvar 1))) A))}~=\\
\texttt{Imp (is\_qexp (CON (Qvar 0)))}\\
\texttt{~(lImp (typeof (CON (Qvar 0)) qubit)}\\
\texttt{~~(Imp (is\_qexp (CON (Qvar 1)))}\\
\texttt{~~~(lImp (typeof (CON (Qvar 1)) qubit)}\\
\texttt{~~~~(atom\_ (typeof~(Prod (CON (Qvar 0)) (CON (Qvar 1))) A)))))}
\end{flushleft}
It is important to note here that both the linear and duplicable
versions of the circuit rule require the subgoal to be proved
intuitionistically. Moreover, the subgoal is the same for both
cases. Although this is in contrast with other rules, it reflects the
real semantics of circuits, because if a circuit has been proven to be
of a certain type, then it should be possible to use as many instances of
this circuit as needed; it is an autonomous component. If it happens
that a free variable appears in a circuit construct, then this variable
is of duplicable type. This semantics is imposed by the above rules
since we keep the list of linear subgoals empty.  According to
\cite{RossPhD15}, quantum variables appearing in the circuit
constructs are called bounded quantum variables. That is why
\texttt{FQ} of a circuit construct is supposed to return an empty
list.
Also, the free variables of \texttt{t} should match the input of the circuit
\texttt{C}, and the free variables of \texttt{a} should match the
output of \texttt{C}. The functions \texttt{circIn} and
\texttt{circOut} are defined as abstract functions. In particular,
\cite{RossPhD15} does not provide specific definitions for these
functions. They are identified only by their domains and
co-domains.
In the Coq code, we define them as variables in a module.  Before
proving properties about a specific set of circuits, these variables
must be instantiated with a particular definition.
Since, like~\cite{RossPhD15}, we are proving only meta-level
properties, we leave them abstract also.

\section{Type Soundness}
\label{sec:sr}
In this section, we formally verify the type soundness of
Proto-Quipper by addressing three important properties: a type
soundness under subtyping rule, inversion lemmas for Proto-Quipper
\emph{values}, and the subject reduction theorem.

\subsection{Context Subtyping}
\label{sec:ctxsub}

In a type system with subtyping, an important soundness lemma is one
that expresses that typing is preserved under the subtype relation
extended to contexts.  In our setting, the statement of this lemma is
in a sense, a general form of the $ax_c$ rule where the contexts in
the premise and conclusion are not the same, but instead have a
subtyping relation between them, in this case a kind of
``contravariant'' one.
The lemma states, roughly, that if
$\mathtt{seq\_~il~ll~(atom\_~A)}$ holds, $\mathtt{A}$ is a subtype of
$\mathtt{B}$, and the pair $\mathtt{(il',ll')}$ is a ``subtype'' of
the pair $\mathtt{(il,ll)}$, then $\mathtt{seq\_~il'~ll'~(atom\_~B)}$
holds.  Context subtyping includes an extension of subtyping in the
obvious way; an atom $\mathtt{typeof~a~t1}$ occurs in $\mathtt{il'}$
or $\mathtt{ll'}$ if and only if an atom $\mathtt{typeof~b~t2}$ occurs
in the corresponding $\mathtt{il}$ or $\mathtt{ll}$ and
$\mathtt{Subtyping~t1~t2}$ holds.

Before we can make the above statement formal, we must consider
additional constraints on the contexts in the sequents.  As discussed
in~\cite{FMP:MSCS17}, in general when formalizing meta-theory,
statements of theorems often relate two or more judgments and if the
contexts in these judgments are non-empty, a \textit{context relation}
is often needed to specify the constraints in the form of a relation
between them.  The above lemma is an example that relates exactly two
statements about the $\mathtt{seq\_}$ predicate with their
corresponding contexts.  A variety of examples in a simpler setting
with an intuitionistic SL are discussed in detail in~\cite{FMP:JAR15}.
We adopt this notion of context relation here, extending it to express
our requirements, which are significantly more complex.  We capture
both the subtyping constraints as well as the necessary additional
constraints in the following inductive definition.

\begin{flushleft}
\texttt{Inductive~Subtypecontext~:=}\\
\texttt{|~subcnxt\_i:~Subtypecontext~[]~[]~[]~[]}\\
\texttt{|~subcnxt\_q:~forall~a~il~il'~ll~ll',}\\
\texttt{~~Subtypecontext~il'~ll'~il~ll~-{>}~}\\
\texttt{~~Subtypecontext~(is\_qexp~a::il')~ll'~(is\_qexp~a::il)~ll}\\
\texttt{|~subcnxt\_iig:~forall~a~t1~t2~il~il'~ll~ll',}\\
\texttt{~~Subtyping~t1~t2~-{>}~(exists~c,~t2~=~bang~c)~-{>}}\\
\texttt{~~Subtypecontext~il'~ll'~il~ll~-{>}}\\
\texttt{~~Subtypecontext~(is\_qexp~a::typeof~a~t1::il')~ll'~}\\
\texttt{~~~~~~~~~~~~~~~~~(is\_qexp~a::typeof~a~t2::il)~ll}\\
\texttt{|~subcnxt\_llg:~forall~a~t1~t2~~il~il'~ll~ll',}\\
\texttt{~~validT~(bang~t1)~-{>}~validT~(bang~t2)~-{>}}\\
\texttt{~~Subtyping~t1~t2~-{>}~Subtypecontext~il'~ll'~il~ll~-{>}}\\
\texttt{~~Subtypecontext~(is\_qexp~a::il')~(typeof~a~t1::ll')~}\\
\texttt{~~~~~~~~~~~~~~~~~(is\_qexp~a::il)~(typeof~a~t2::ll)}\\
\texttt{|~subcnxt\_lig:~forall~a~t1~t2~il~il'~ll~ll',}\\
\texttt{~~validT~(bang~t2)~-{>}~(exists~c,~t1~=~bang~c)~-{>}}\\
\texttt{~~Subtyping~t1~t2~-{>}~Subtypecontext~il'~ll'~il~ll~-{>}}\\
\texttt{~~Subtypecontext~(is\_qexp~a::typeof~a~t1::il')~ll'}\\
\texttt{~~~~~~~~~~~~~~~~~(is\_qexp~a::il)~(typeof~a~t2::ll).}
\end{flushleft}
The additional constraints include, for example, that every time there
is a typing assumption $\mathtt{(typeof~a~t)}$ in either a linear or
intuitionistic context, there must be an assumption of the form
$\mathtt{(is\_qexp~a})$ in the intuitionistic context.  Note that we can
write $\mathtt{Subtypecontext~il~ll~il~ll}$, where the first and third
arguments are the same, and similarly for the second and fourth, when we
want to ignore subtyping and care only that these additional
constraints are met.  We will use $\mathtt{Subtypecontext}$ for this
secondary purpose rather than define a new context relation.

Before expressing and proving the central lemma mentioned above, we
have to tackle a crucial theorem that is very helpful for splitting a linear context:

\begin{flushleft}
\texttt{Theorem~subcnxt\_split:~forall~il~il'~ll~ll'~ll1~ll2,}\\
\texttt{~~Subtypecontext~il'~ll'~il~ll~-{>}~split~ll~ll1~ll2~->}\\
\texttt{~~exists~il1~il2~ll1'~ll2',}\\
\texttt{~~~~split~ll'~ll1'~ll2'~/{\char'134}}\\
\texttt{~~~~(forall~a,~In~a~il~-{>}~In~a~il1)~/{\char'134}~}\\
\texttt{~~~~(forall~a,~In~a~il~-{>}~In~a~il2)~/{\char'134}~}\\
\texttt{~~~~Subtypecontext~il'~ll1'~il1~ll1~/{\char'134}}\\
\texttt{~~~~Subtypecontext~il'~ll2'~il2~ll2.}\\
\end{flushleft}
\noindent This theorem explains the effect of splitting a linear
context in the \texttt{Sub\-type\-con\-text} relation. Recall that list
splitting means dividing a list into two lists with elements in any
order. This theorem helps when we are dealing with several subgoals in
\texttt{splitseq} or multiplicative conjunction goals, and we want to split the linear context without
losing the benefit of the \texttt{Subtypecontext} relation. This situation is pretty common in our proofs.

We are now ready to state the central lemma.
\begin{flushleft}
\texttt{Theorem~subtypecontext\_subtyping:~forall~a~IL~IL'~LL~LL'~B~A,~}\\
\texttt{~~Subtypecontext~IL'~LL'~IL~LL~-{>}~}\\
\texttt{~~seq\_~IL~LL~(atom\_~(typeof~a~A))~-{>}~Subtyping~A~B~-{>}}\\
\texttt{~~seq\_~IL'~LL'~(atom\_~(typeof~a~B)).}\\
\end{flushleft}

\noindent
With the help of the above lemma and others, we successfully prove
this theorem.
The proof amounts to 800 lines of Coq script. The proof is by
induction on the height of the sequent derivations, with
cases for every possible typing rule defined in the program context
\texttt{prog}.

\subsection{Inversion Rules for Values}\label{sec:invrules}

A Proto-Quipper expression is considered a \emph{value}, or
non-reducible expression,  if it matches one of the following cases:
\begin{flushleft}
\texttt{Inductive~is\_value:~qexp~-{>}~Prop~:=~}\\
\texttt{~~Varv:~forall~x,~is\_value~(Var~x)}\\
\texttt{|~Qvarv:~forall~x,~is\_value~(CON~(Qvar~x))}\\
\texttt{|~Circv:~forall~a~t~i,~quantum\_data~t~-{>}~quantum\_data~a~-{>}}\\
\texttt{~~~~~~~~~is\_value~(Circ~t~i~a)}\\
\texttt{|~Truev:~is\_value~(CON~TRUE)} \texttt{~~~|~Falsev:~is\_value~(CON~FALSE)}\\
\texttt{|~Starv:~is\_value~(CON~STAR)} \texttt{~~~|~Boxv:~is\_value~(CON~BOX)}\\
\texttt{|~Unboxv:~is\_value~(CON~UNBOX)} \texttt{~|~Revv:~is\_value~(CON~REV)}\\
\texttt{|~Funvv:~forall~f,~abstr~f~-{>}~is\_value~(Fun~f)}\\
\texttt{|~Prodv:~forall~v~w,~is\_value~v~-{>}~is\_value~w~-{>}}\\
\texttt{~~~~~~~~~is\_value~(Prod~v~w)}\\
\texttt{|~Unboxappv:~forall~v,~is\_value~v~-{>}}\\
\texttt{~~~~~~~~~~~~~is\_value~(App~(CON~UNBOX)~v).}\\
\end{flushleft}

\noindent The major role of these special expressions is  in defining the language reduction
rules (i.e., operational semantics), as detailed in the next section, where the main objective is to reduce
a Proto-Quipper expression to one of these forms. The cases mentioned above in the definition
are obvious, except for \texttt{Unboxappv}. The reason behind considering
\texttt{(App~(CON~UNBOX)~v)} a \emph{value} is because the resulting
expression is of function type; as stated in~\cite{RossPhD15}, the unbox
operator turns a circuit into a circuit-generating function, and thus
this case is similar to \texttt{Funvv}.

Now that we have the language \emph{values}, we can prove a number of
inversion lemmas (corresponding to lemmas in~\cite{RossPhD15}).
In each, we prove that a well typed value
\texttt{v} should follow certain Proto-Quipper format(s). Here is the
first example:
\begin{flushleft}
\texttt{Theorem~sub\_one\_inv:~forall~IL~a,~}\\
\texttt{~~{\char'176}(In~(is\_qexp~(CON~UNBOX))~IL)~-{>}~is\_value~a~-{>}~}\\
\texttt{~~Subtypecontext~IL~[]~IL~[]~-{>}~{\char'176}(In~(is\_qexp~a)~IL)~-{>}}\\
\texttt{~~seq\_~IL~[]~(atom\_~(typeof~a~one))~-{>}~a~=~CON~STAR.}\\
\end{flushleft}

\noindent In this particular inversion theorem, a value of type
\texttt{one} should be the \texttt{STAR} constant. To make sure that the
context is not misused, we must prevent assumptions about
well-formedness of value terms from occurring in the context. To do
so, we use the \texttt{Subtypecontext} relation
to ensure that every \texttt{(is\_qexp a)} that appears in
\texttt{IL} is associated with a typing atom \texttt{(typeof a A)} in
\texttt{IL}
This way, if
\texttt{{\char'176}(In~(is\_qexp~a)~IL)} is assumed in the inversion
theorem then no typing hypothesis for the value ``\texttt{a}'' appears in
\texttt{IL}.
We add a similar condition for
the \texttt{UNBOX} due to the fact that \texttt{(App~(CON~UNBOX)~v)}
is a value: if \texttt{UNBOX} is assumed in \texttt{IL} to be of type
\texttt{(arrow A one)} (which is a false assumption according to
Proto-Quipper typing rules), then by assuming the existence of a value
\texttt{v} of type \texttt{A}, one can prove that ``\texttt{a}'' has the
form \texttt{\texttt{(App~(CON~UNBOX)~v)}}, which is wrong. The
conclusion of the above theorem is also valid for the type
\texttt{(bang one)}, which is also formally proved. Similar results
have been proved for all other values. To avoid repetition, we pick
one more interesting example to illustrate, however, the reader still
can find a full list of the inversion rules online at
\cite{ourformal}.

The following theorem is concerned with the values of the
\texttt{arrow} type constructor. In contrast to the previous theorem
(and all other cases), we have several possibilities that satisfy this
typing format:
\begin{flushleft}
\texttt{Theorem~sub\_bangarrow\_inv:~forall~IL~LL~a~T~U,~}\\
\texttt{~~{\char'176}(In~(is\_qexp~(CON~UNBOX))~IL)~-{>}~valid~T~-{>}~valid~U~-{>}}\\
\texttt{~~(forall v, a = App~(CON~UNBOX) v~-{>}~{\char'176}(In~(is\_qexp~v)~IL))~-{>}}\\
\texttt{~~is\_value~a~-{>}~Subtypecontext~IL~LL~IL~LL~-{>}}\\
\texttt{~~{\char'176}(In~(is\_qexp~a)~IL)~-{>}}\\
\texttt{~~seq\_~IL~LL~(atom\_~(typeof~a~(bang~(arrow~T~U))))~-{>}}\\
\texttt{~~(exists~f,~a~=~Fun~f)~{\char'134}/~(exists T0,~(a~=~CON~(BOX T0))~{\char'134}/}\\
\texttt{~~(a~=~CON~UNBOX)~{\char'134}/~(a~=~CON~REV)~{\char'134}/~}\\
\texttt{~~(exists~t~i~u,~a~=~App~(CON~UNBOX)~(Circ~t~i~u)).}\\
\end{flushleft}

\noindent According to above theorem, the value ``\texttt{a}'' would be a function, a
quantum circuit conversion function, or a function application over the \texttt{UNBOX} constant.

The proofs of the above inversion rules are done by case analysis of
the possible values. For each case, it is proved that
\texttt{seq\_~IL~LL~(atom\_~(typeof~a~A))} requires that
\texttt{(In~(typeof~a~A)~IL)} or \texttt{(In~(typeof~a~A)~LL)}, which
leads to a contradiction except for the particular value we are
interested in. For instance, in the theorem
\texttt{sub\_one\_inv}, assuming that
\texttt{seq\_ IL LL (atom\_ (typeof a one))} leads to a contradiction
in all cases except for the \texttt{STAR} constant.

We also prove inversion rules that serve the opposite purpose from the
above ones, where the expression format is provided and the theorems
conclude the right corresponding type and/or typing rules. Those rules
are more general, i.e., they are not restricted to values. For
instance, the following theorem characterizes the typing rules for the
\texttt{Slet} statement:
\begin{flushleft}
\texttt{Theorem~sub\_slet\_inv:~forall~IL~LL~a~b~A,}\\
\texttt{~~Subtypecontext~IL~LL~IL~LL~-{>}~}\\
\texttt{~~seq\_~IL~LL~(atom\_~(typeof~(Slet~a~b)~A))~-{>}}\\
\texttt{~~{\char'176}(In~(is\_qexp~(Slet~a~b))~IL)~-{>}}\\
\texttt{~~(exists~B,~Subtyping~B~A~/{\char'134}~~}\\
\texttt{~~~(splitseq\_~IL~LL}\\
\texttt{~~~~~[Conj~(atom\_~(typeof~a~B))~(atom\_~(typeof~b~(bang~one)))]~{\char'134}/}\\
\texttt{~~~~splitseq\_~IL~LL~}\\
\texttt{~~~~~[Conj~(atom\_~(typeof~a~B))~(atom\_~(typeof~b~~one))]))~{\char'134}/}\\
\texttt{~~(validT~A~/{\char'134}~~~}\\
\texttt{~~~(splitseq\_~IL~LL}\\
\texttt{~~~~~[Conj~(atom\_~(typeof~a~A))~(atom\_~(typeof~b~(bang~one)))]~{\char'134}/}\\
\texttt{~~~~splitseq\_~~IL~LL}\\
\texttt{~~~~~[Conj~(atom\_~(typeof~a~A))~(atom\_~(typeof~b~one))])).}
\end{flushleft}

\noindent The \texttt{Subtypecontext} is used here again to ensure
that we are dealing with a valid typing context as explained earlier;
it has nothing to do with the subtyping between contexts since the
super-context and subcontext are the same. The above theorem concludes
two possible cases for a well-typed \texttt{Slet} expression.  First,
the right side of the main disjunction corresponds to the case when
the type \texttt{A} is deduced directly from the \texttt{SLet} typing
rules (see the \texttt{tsletl} and \texttt{tsleti} of the
\texttt{Prog} definition in \cite{ourformal}).  Second, the left
disjunct corresponds to the case when the typing deduction has gone
through one or more subtyping rules (see rules \texttt{axc1} and
\texttt{axc2} of the \texttt{prog} definition discussed above and in
\cite{ourformal}) followed by the regular \texttt{SLet} typing rules).

Here is another example for the circuit construct:

\begin{flushleft}
\texttt{Theorem~sub\_Circ\_inv:~forall~IL~LL~t~a~c~A,}\\
\texttt{~~Subtypecontext~IL~LL~IL~LL~-{>}~}\\
\texttt{~~seq\_~IL~LL~(atom\_~(typeof~(Circ~t~c~a)~A))~-{>}}\\
\texttt{~~{\char'176}(In~(is\_qexp~(Circ~t~c~a))~IL)~-{>}}\\
\texttt{~~(exists~T~T'~B,}\\
\texttt{~~~~Subtyping~B~A~/{\char'134}~validT~(circ~T~T')~/{\char'134}~LL~=~[]~/{\char'134}}\\
\texttt{~~~~splitseq\_~IL~[]~}\\
\texttt{~~~~~[And~(toimp~(FQ~a)~(atom\_(typeof~a~T')))}\\
\texttt{~~~~~~~~~~(toimp~(FQ~t)~(atom\_(typeof~t~T)))]~/{\char'134}}\\
\texttt{~~~~(B~=~circ~T~T'~{\char'134}/~B~=~bang~(circ~T~T')))~{\char'134}/}\\
\texttt{~~(exists~T~T',}\\
\texttt{~~~~validT~(circ~T~T')~/{\char'134}~LL~=~[]~/{\char'134}}\\
\texttt{~~~~splitseq\_~IL~[]}\\
\texttt{~~~~~[And~(toimp~(FQ~a)~(atom\_(typeof~a~T')))}\\
\texttt{~~~~~~~~~~(toimp~(FQ~t)~(atom\_(typeof~t~T)))]~/{\char'134}}\\
\texttt{~~~~(A~=~circ~T~T'~{\char'134}/~A~=~bang~(circ~T~T'))).}\\
\end{flushleft}

\noindent An interesting fact about a well-typed circuit construct is
that its linear context should be
empty, and this is expected as the circuit is supposed to be used
several times, and hence it does not depend on linear typing
atoms. Recall that the  quantum variables that appear in ``\texttt{t}''
and ``\texttt{a}'' are considered bound variables.

Similar results have been proved for other Proto-Quipper expressions, e.g., functions, function application, circuit conversion constants, etc.

\subsection{Subject Reduction}
Reduction rules, i.e., operational semantics, are crucial for any
programming language, defining how valid expressions in a language are
simplified until a non-reducible expression is reached, i.e.  a value.
In the context of a language's reduction rules, it is very important
to ensure the integrity of typing rules by ensuring that a reduction
of a well-typed expression should not affect the expression's type
(i.e., type soundess). In Proto-Quipper, there are seventeen reduction
rules, which handle all possible cases while ensuring that only one
reduction rule can be applied at a time.  To keep the description
concise, we choose five reduction rules to explain in this paper, which
are good representative samples.
Similar to typing rules, we encode Proto-Quipper reduction rules as
part of the inductive definition of \texttt{prog}.
For the complete set of rules, we refer the reader to
\cite{RossPhD15}, and for their complete encoding in Coq, see
\cite{ourformal}.

Given a  closure  pair $[C, a]$, where $C$ is a circuit constant and
$a$ is a Proto-Quipper expression  such that $FQ(a) \in Out(C)$, an
\texttt{if} statement is reduced according to the rules listed in Figure
\ref{tab:ifrr}.
\begin{figure}
\centering
  \begin{tabular}{c}
\AxiomC{$[C, a]\rightarrow[C', a']$}
\RightLabel{$cond$}
\UnaryInfC{$[C, if~a~then~b~else~c]\rightarrow[C', if~a'~then~b~else~c]$}
\DisplayProof
\\ \\
\AxiomC{}
\RightLabel{$ifT$}
\UnaryInfC{$[C, if~True~then~b~else~c]\rightarrow[C, b]$}
\DisplayProof
\\ \\

\AxiomC{}
\RightLabel{$ifF$}
\UnaryInfC{$[C, if~False~then~b~else~c]\rightarrow[C, c]$}
\DisplayProof

 \end{tabular}
\caption{If statement reduction rules}
\label{tab:ifrr}
\end{figure}
\noindent The formal Coq presentation of the rules in Figure
\ref{tab:ifrr} as part of the \texttt{prog} definition is as follows:
\begin{flushleft}
\texttt{|~ifr:~forall~C~C'~b~b'~a1~a2,~valid\_c~C~(If~b~a1~a2)~->}\\
\texttt{~~valid\_c~C'~(If~b'~a1~a2)~-{>}~{\char'176}(is\_value~b)~-{>}}\\
\texttt{~~prog~(reduct~C~(If~b~a1~a2)~C'~(If~b'~a1~a2))}\\
\texttt{~~~~[atom\_~(reduct~C~b~C'~b');}\\
\texttt{~~~~~atom\_~(is\_qexp~a1);~atom\_~(is\_qexp~a2)]~[]}\\
\texttt{|~truer:~forall~C~a1~a2,~valid\_c~C~(If~(CON~TRUE)~a1~a2)~->}\\
\texttt{~~prog~(reduct~C~(If~(CON~TRUE)~a1~a2)~C~a1)}\\
\texttt{~~~~[atom\_~(is\_qexp~a1);~atom\_(is\_qexp~a2)]~[]}\\
\texttt{|~falser:~forall~C~a1~a2,~valid\_c~C~(If~(CON~FALSE)~a1~a2)~->}\\
\texttt{~~prog~(reduct~C~(If~(CON~FALSE)~a1~a2)~C~a1)}\\
\texttt{~~~~[atom\_~(is\_qexp~a1);~atom\_(is\_qexp~a2)]~[]}\\
\end{flushleft}

\noindent where \texttt{valid\_c} ensures that
[\texttt{C',(If~b'~a1~a2)}] forms a valid closure. Note that the rule
\texttt{ifr} is only applicable if ``\texttt{b}'' is not a value,
otherwise this rule would be applied an infinite number of times
without achieving any progress. When ``\texttt{b}'' is a value, i.e.,
$True$ or $False$ one of the other two rules can be applied.

The remaining examples of Proto-Quipper reduction rules that we
consider here are the circuit construct and its conversion function
boxing and unboxing rules, which appear in Figure \ref{tab:boxing}.
\begin{figure}
\centering
  \begin{tabular}{c}
\AxiomC{$[D, a]\rightarrow[D', a']$}
\RightLabel{$circ$}
\UnaryInfC{$[C, (t,D,a)]\rightarrow[C, (t,D',a')]$}
\DisplayProof
\\ \\
\AxiomC{$Spec_{FQ(v)}(T) = t$}
\AxiomC{$New(FQ(t)) = D$}
\RightLabel{$box$}
\BinaryInfC{$[C, box (v)]\rightarrow[C, (t,D,v~t)]$}
\DisplayProof

\\ \\
\AxiomC{$bind(u,v) = b$}
\AxiomC{$Append(C,D,b) = (C',b')$}
\RightLabel{$unbox$}
\BinaryInfC{$[C, (unbox~(u,D,u'))~v]\rightarrow[C', b'(u')]$}
\DisplayProof
 \end{tabular}
  \caption{Circuit reduction rules}
\label{tab:boxing}
\end{figure}
${Spec}_Q$ is a function that given a quantum data type $T$, returns a
\emph{specimen} $t$ of that type, i.e., a quantum data term having
type $T$ that has the further property that all quantum variables in $t$
are ``fresh'' with respect to the quantum variables occuring in $Q$.
${New}$ is a function that creates a new identity
circuit, i.e., given a set of quantum variables, these variables serve
as the inputs and outputs, and the new circuit contains only wires
where inputs are mapped directly to the same outputs.

The notion of a \emph{binding} expresses the way in which wires should
be connected.  The $\mathit{bind}$ function is a bijection on quantum
variables, where it wires each quantum variable in $u$ with the
corresponding one in $v$.  For example, $bind(\langle q_1,q_2\rangle,
\langle q_3,q_4\rangle) = \{(q_1,q_3);(q_2,q_4)\}$. The $\mathit{Append}$
function is responsible for appending two circuits. It returns the new
circuit and a binder to rename the outputs of the newly created
circuit. This rename step is not really required since the newly
created circuit can use the same names of the output of the circuit
$D$ as its output; however, it is considered an added feature in the
Proto-Quipper language. The following are the corresponding formal
reduction rules of the $box$ and $unbox$ rules listed in Figure
\ref{tab:boxing}:
\begin{flushleft}
\texttt{|~boxr:~forall~v~T~t~C,~valid~T~-{>}~is\_value~v~->}\\
\texttt{~~t~=~(Spec~(newqvar~v)~T)~->}\\
\texttt{~~prog~(reduct~C~(App~(CON~(BOX~T))~v)~C~}\\
\texttt{~~~~~~~~~(Circ~t~(circNew~(FQ~t))~(App~v~t)))}\\
\texttt{~~~~[atom\_~(is\_qexp~~v)]~[]}\\
\texttt{|~unboxr:~forall~u~u'~v~C~C'~D~b',}\\
\texttt{~~quantum\_data~u~-{>}~quantum\_data~u'~-{>}~quantum\_data~v~-{>}}\\
\texttt{~~bind~u~v~{->}~bind~u'~b'~{->} }\\
\texttt{~~circApp~(Crcons~C)~(Crcons~D)~=~C'~->}\\
\texttt{~~prog~(reduct~(Crcons~C)}\\
\texttt{~~~~~~~~~~~~~~~(App~(App~(CON~UNBOX)~(Circ~u~D~u'))~v)~C'~b')~}\\
\texttt{~~~~[atom\_~(is\_qexp~v);~atom\_~(is\_qexp~b')]~[]}\\
\end{flushleft}
In the $\mathtt{boxr}$ clause, the function $\mathtt{newqvar}$ returns
a variable that is fresh with respect to its argument, which is a
quantum term.  In our implementation of the $\mathtt{Spec}$ function,
the first argument is a natural number and all quantum variables in
the result are represented by numbers greater than this input number.
Together $\mathtt{Spec}$ and $\mathtt{newqvar}$ implement the $Spec$
function in the $box$ rule.

In our formalization, we implement a relation $\mathtt{(bind~u~v)}$,
which holds if $\mathtt{u}$ and $\mathtt{v}$ are related by some
binding, i.e., they are the same quantum data terms up to the renaming
of quantum variables.  This relation is used twice in the
$\mathtt{unboxr}$ clause to implement the binding relations in the
$unbox$ rule.
We note that the $\mathit{Append}$ function is formalized as an abstract function without concrete implementation, where Proto-Quipper does not provide one; rather it provides its properties which we axiomatize in the above rule.

Similar rules have been formalized for the rest of Proto-Quipper
expressions and can be found in \cite{ourformal}.

Now, we can present the most important property of the Proto-Quipper
language, namely subject reduction, which completes the type soundness result
of the Proto-Quipper type system and its operational semantics:

\begin{flushleft}

\texttt{Theorem~subject\_reduction:~forall~i~IL~C~C'~a~a'~LL1~LL2~A,}\\
\texttt{~~(forall~V,~In~V~(get\_boxed~a)~-{>}}\\
\texttt{~~~~{\char'176}(exists~n,~V = (Var~n)~{\char'134}/~V = (CON~(Qvar~n))))~-{>}}\\
\texttt{~~NoDup~(FQUC~a')~-{>}~NoDup~(FQU~a')~-{>}}\\
\texttt{~~(forall~t,~In~t~IL~-{>}}\\
\texttt{~~~~(exists~ n,~t~=~is\_qexp~(Var~n)~{\char'134}/}\\
\texttt{~~~~~~~~~~~~~~~~t~=~is\_qexp~(CON~(Qvar~n))~{\char'134}/}\\
\texttt{~~~~~~~~~~~~~~~~exists~T,~t = typeof~(Var~n)~T))~-{>}}\\
\texttt{~~(forall~q,~In~q~(FQ~a')~-{>}~In~(typeof~q~qubit)~LL2)~-{>}}\\
\texttt{~~Subtypecontext~IL~LL1~IL~LL1~-{>}}\\
\texttt{~~Subtypecontext~IL~LL2~IL~LL2~-{>}}\\
\texttt{~~common\_ll~a~a'~LL1~LL2~-{>}}\\
\texttt{~~\underline{seq\_~i~IL~[]~(atom\_~(reduct~C~a~C'~a'))}~-{>}}\\
\texttt{~~\underline{exists~j,~seq\_~j~IL~LL1~(atom\_~(typeof~a~A))}~-{>}}\\
\texttt{~~\underline{exists~k,~seq\_~k~IL~LL~(atom\_~(typeof~a'~A))}.}\\
\end{flushleft}
\noindent The three underlined predicates represent the core of the the subject reduction theorem. The three predicates translate the typical
meaning of the subject reduction: \emph{an expression \texttt{a'}, that is the reduction of a well-typed  expression \texttt{a}, is also well-typed and hold the same type as the original expression \texttt{a}}.

The remaining hypotheses belong to two categories and require
explanation.  The three lines following the theorem name are
additional well-formedness constraints.  As we will describe below,
the one involving \texttt{get\_boxed} was unexpected and illustrates
the merits of formal proofs to check paper-and-pencil ones, which
often overlook the kind of detail captured by this definition.  The
remaining lines are conditions on contexts.  We are able to reuse our
\texttt{Subtypecontext} definition to capture some of the
requirements, but in addition, we define an additional context
relation \texttt{common\_ll} as well as add some additional
requirements on the form of assumptions appearing in contexts.

In general, type soundness results are often stated and proved for
typing of closed terms and empty contexts.  Here, we could restrict
terms so that they contain no free term variables (in particular,
require that \texttt{a} and \texttt{a'} contain no free term
variables), but we cannot do the same for quantum variables since
there is no lambda binder for quantum variables in the language.  Most
of our additional hypotheses provide the required structure for the
appearance of assumptions about quantum variables in contexts, but we
go beyond that and also allow free term variables in \texttt{a} and
\texttt{a'}.  Doing so requires some additional hypotheses in the
statement of the theorem, but allows us to prove a more general form.
The version requiring \texttt{a} and \texttt{a'} to be closed follows
as a corollary.  In summary, the basic requirements for free quantum
variables are that for every variable \texttt{q} (of the form
\texttt{(CON~(Qvar~n))}), \texttt{q} is free in \texttt{a} or
\texttt{a'} if and only if there is an assumption \texttt{(is\_qexp
  q)} in \texttt{IL}, \texttt{q} is free in \texttt{a} if and only if
there is an assumption of the form \texttt{(typeof q qubit)} in
\texttt{LL1}, and \texttt{q} is free in \texttt{a'} if and only if there
is an assumption of the form \texttt{(typeof q qubit)} in
\texttt{LL2}.

We explain the hypotheses in more detail one by one in the following.

\begin{itemize}
\item
\begin{flushleft}
\texttt{(forall~t,~In~t~IL~-{>}}\\
\texttt{~~(exists~ n,~t~=~is\_qexp~(Var~n)~{\char'134}/}\\
\texttt{~~~~~~~~~~~~~~t~=~is\_qexp~(CON~(Qvar~n))~{\char'134}/}\\
\texttt{~~~~~~~~~~~~~~exists~T,~t = typeof~(Var~n)~T))}:\\
\end{flushleft}

This hypothesis restricts the intuitionistic context so that it
includes only typing and well-formedness assumptions about term
variables and quantum variables.  Note that the inversion rules
\texttt{sub\_one\_inv}, \texttt{sub\_bangarrow\_inv},
\texttt{sub\_slet\_inv}, and \texttt{sub\_Circ\_inv} in
Section \ref{sec:invrules} all restrict the form of the argument to
\texttt{is\_qexp}, which is the expression subject to inversion, when
it appears in the intuitionistic context.  In these theorems, the
restrictions are carried over to the linear contexts because of the
way the contexts are related according to the \texttt{Subtypecontext}
predicate.  The same applies here.

\item \texttt{Subtypecontext~IL~LL1~IL~LL1} and
  \texttt{Subtypecontext~IL~LL2~IL~LL2}:

These predicates ensures that the constraints discussed in
Section~\ref{sec:ctxsub} about the relationship between the
intuitionistic and linear contexts are met.  For example, each typing
assumption of the form \texttt{(typeof a t)} in any context must be
associated with a well-formedness assumption \texttt{(is\_qexp a)} in
the intuitionistic context.  In addition, it restricts the linear
context to contain \texttt{typeof} assumptions only.

\item \texttt{common\_ll~a~a'~LL1~LL2}:

One might wonder why the theorem requires two different linear
contexts (\texttt{LL1} and \texttt{LL2}); in particular, the common
subject reduction for classical OLs maintains the same context in all
judgments. The answer is \emph{quantum variables}. Proto-Quipper
reduction rules allow for quantum variable renaming (i.e., in the case
of circuit appending), and allow introduction of new quantum variables
(e.g., a circuit with zero inputs that produces quantum variables at
the output, which typically represents an initialization circuit).  As
a result, the linear context of the original expression is usually a
bit different from the context for the reduced one; they are similar
modulo quantum variables. That is why we have defined the predicate
\texttt{common\_ll}, which ensures that all typing assumptions in
\texttt{LL1} and \texttt{LL2} are the same except for quantum
variables, where quantum variables in \texttt{LL1} belong to the set
of free quantum variables of \texttt{a}, and quantum variables in
\texttt{LL2} belong to the set of free quantum variables of
\texttt{a'}.

\item \texttt{(forall~q,~In~q~(FQ~a')~-{>}~In~(typeof~q~qubit)~LL2)}:

The previous predicate does not assure that \texttt{LL2} contains
assumptions about all free quantum variables of \texttt{a'}; it only
ensures if a quantum variables exists in \texttt{LL2} then it belongs
to \texttt{(FQ~a')}. That is why we added this assumption. One might
ask why not to do the same for \texttt{LL1}; the answer that we do not
need that assumption since \texttt{a} is well-typed under \texttt{LL1}
and hence it must contain all quantum variable of
\texttt{(FQ~a)}. This fact is proved as per the following theorem
(note the equivalence in the last two lines):
\begin{flushleft}
\texttt{Theorem LL\_FQ:~forall~u~A~IL~LL,}\\
\texttt{~~(forall~q~T,}\\
\texttt{~~~~In~(typeof (CON (Qvar q)) T) LL -{>} T = qubit) -{>}}\\
\texttt{~~(forall q T, In (typeof (CON (Qvar q)) T) LL -{>}}\\
\texttt{~~~~count\_occ  eq\_dec LL (typeof (CON (Qvar q)) T) = 1) -{>}}\\
\texttt{~~(forall~t,~In~t~IL~-{>}}\\
\texttt{~~~~(exists~ n,~t~=~is\_qexp~(Var~n)~{\char'134}/}\\
\texttt{~~~~~~~~~~~~~~~~t~=~is\_qexp~(CON~(Qvar~n))~{\char'134}/}\\
\texttt{~~~~~~~~~~~~~~~~exists~T,~t = typeof~(Var~n)~T))~-{>}}\\
\texttt{~~Subtypecontext IL LL IL LL -{>}}\\
\texttt{~~seq\_ IL LL (atom\_ (typeof u A)) -{>}}\\
\texttt{~~\textbf{(forall q, In (CON (Qvar q)) (FQ u) <->}}\\
\texttt{~~~~~~~~~~~~~\textbf{In (typeof (CON (Qvar q)) qubit) LL)}}.
\end{flushleft}

\item \texttt{NoDup~(FQUC~a')} and \texttt{NoDup~(FQU~a')}:

\texttt{FQU} is similar to \texttt{FQ} in that it returns the free
quantum variables occurring in a term.  The difference is that
\texttt{FQ} returns a set, while \texttt{FQU} returns a list so that
if there is more than one occurrence of a free variable in a term, it
occurs more than once in the output list. \texttt{FQUC} is similar,
but also includes bounded quantum variables of quantum circuits.  (In
\texttt{FQ} and \texttt{FQU}, those variables are excluded.)  The
predicate \texttt{NoDup} comes from Coq's list library and is true of
lists that contain no duplicates.  In quantum calculus, quantum
variables are processed once. We use \texttt{FQU} to rule out
expressions that involve quantum replicas. \texttt{FQUC} ensures no
quantum variable replicas inside circuit bodies.

\item
\begin{flushleft}
\texttt{(forall~V,~In~V~(get\_boxed~a)~-{>}}\\
\texttt{~~{\char'176}(exists~n,~V = (Var~n)~{\char'134}/~V = (CON~(Qvar~n))))}:
\end{flushleft}

The function \texttt{get\_boxed} returns a list of all expression that
are arguments to the $box^T$ operator, i.e., for the expression
\texttt{(APP (CON (BOX T)) a)}, it returns \texttt{a}. We want to make
sure that those expressions are neither free variables nor quantum
variables. Unlike the other hypotheses in the theorem, which are
mostly due to our method of formalization, this condition should be
respected in the language implementation itself; otherwise the
language allows for expressions that are not \emph{values} and there
is no reduction rule that applies.  Without this, neither subject
reduction nor progress properties hold for Proto-Quipper.
\end{itemize}

The proof of the above theorem involves several lemmas and theorems
and makes significant use of a number of the inversion rules discussed
in the previous section.  It is proved by induction on \texttt{i}, the
height of the proof of the sequent containing the typing judgment,
followed by a case analysis of 17 reduction rules, each of which
require at least 10 major subgoals due to case analysis and
inversions.

This concludes our formalization of Proto-Quipper in Hybrid, along
with our development of a generic linear specification logic used as a
platform to reason about quantum programming languages. In
the following section, we will discuss a very critical aspect of our
development where we ensure that the developed formalization models
the intended behavior of Proto-Quipper as presented in
\cite{RossPhD15}.

\section{Adequacy}
\label{sec:adequacy}

It is important to show that both syntax and inference rules are
adequately represented in Hybrid.  Adequacy of syntax encoding, also
called \textit{representational adequacy}, is discussed for the
lambda calculus as an OL in Hybrid in~\cite{AmblerEtAl:TPHOLs02}
and proved in detail in~\cite{Crole:MSCS11}, while adequacy for a
fragment of a functional programming language known as Mini-ML is
proved in~\cite{FeltyMomigliano:JAR12}. Proto-Quipper contains the
lambda calculus as a sublanguage, and representational adequacy for
the full language is a straightforward extension of these other
results. We give a brief overview here.

First, the $\mathtt{proper}$ and $\mathtt{abstr}$ predicates, defined
in the Hybrid library, provide important tools for proving
representational adequacy of any OL.  As mentioned, the
$\mathtt{proper}$ predicate rules out terms with dangling indices,
which clearly cannot occur in well-formed terms of any OL.  The
$\mathtt{abstr}$ predicate, as mentioned, applies to arguments of the
$\mathtt{lambda}$ operator.  Recall that this operator has one
argument of functional type $(\mathtt{expr}\rightarrow\mathtt{expr})$.
To adequately represent OL syntax, we must rule out \emph{exotic}
functions, i.e., functions that do not encode OL lambda terms.  To
illustrate, we consider the example from~\cite{FeltyMomigliano:JAR12}.
Suppose we have $(\mathtt{lambda}~e)$, where $e = (\lambda
x.\mathtt{count}~x)$ where $(\mathtt{count}~x)$ counts the total
number of variables and constants occurring in $x$.  This function
clearly does not represent syntax.  Only functions that behave
\emph{uniformly} or \emph{parametrically} on their arguments, such as
the one discussed in the introduction, represent OL terms.  The
$\mathtt{abstr}$ predicate identifies exactly this set of functions.
See~\cite{DFHtlca95} and the related work section
of~\cite{FeltyMomigliano:JAR12} for a careful analysis of this
phenomenon.  The former applies specifically in the context of Coq.
Also, see~\cite{ourformal} for the precise definitions of these two
predicates.

Second, proving representational adequacy requires defining an
encoding function between OL terms and their representation in Hybrid,
and showing that this function is a bijection.  In particular, we
write $\encode{\Gamma}{\cdot}$ for the encoding function from
Proto-Quipper terms with free term variables in $\Gamma$ to terms of
type \texttt{qexp}.  We omit its definition, but note that it maps
each term variable $x$ in $\Gamma$ to a distinct Coq variable
$\mathtt{x:qexp}$, and each quantum variable $q$ in $\Gamma$ to an
expression of the form $\mathtt{(CON~(Qvar~}q_i\mathtt{))}$, where
$q_i$ is a distinct natural number for each quantum variable.
Consider the judgment $\Gamma \vdash
\mathit{is\_qexp}~a$ and a full set of inference rules for this
judgment in the style of those presented in
Section~\ref{sec:encodepqsyntax}. Let $\{x_1,\ldots,x_n\}$ be the set
of term variables in $\Gamma$ and let
$\mathtt{x_1},\ldots,\mathtt{x_n}$ be the encodings of these
variables.  We must prove that for any Proto-Quipper term $a$, if
$\Gamma \vdash \mathit{is\_qexp}~{a}$, then the following is provable
in Coq:
$$\begin{array}{l}
  \mathtt{proper~x_1}\rightarrow\cdots\mathtt{proper~x_n}\rightarrow{}\\
  \mathtt{exists~(i:nat),}\\
  \mathtt{~~seq~i~[(is\_qexp~x_1);\ldots;(is\_qexp~x_n)]~[]}\\
  \mathtt{~~~~~~~~(atom\_~(is\_qexp}~\encode{\Gamma}{a}))
\end{array}$$
Furthermore, we write $\decode{\mathtt{\Gamma}}{\cdot}$ for the
decoding function. Let $\mathtt{\Gamma}$ be Coq variables and terms
$\{\mathtt{x_1}, \ldots, \mathtt{x_n},
\mathtt{(CON~(Qvar~}q_{i_1}\mathtt{))}, \ldots,
\mathtt{(CON~(Qvar~}q_{i_m}\mathtt{))}\}$ of type \texttt{qexp}, let
$\mathtt{E}$ be a term of type \texttt{qexp}, and let $\Gamma$ be:
$$\{\decode{\mathtt{\Gamma}}{\mathtt{x_1}}, \ldots,
\decode{\mathtt{\Gamma}}{\mathtt{x_n}},
\decode{\mathtt{\Gamma}}{\mathtt{CON~(Qvar~}q_{i_1}\mathtt{)}}, \ldots,
\decode{\mathtt{\Gamma}}{\mathtt{CON~(Qvar~}q_{i_m}\mathtt{)}}\}.$$
We must prove that if the following is provable in Coq:
$$\begin{array}{l}
  \mathtt{proper~x_1}\rightarrow\cdots\mathtt{proper~x_n}\rightarrow{}\\
  \mathtt{exists~(i:nat),}\\
  \mathtt{~~seq~i~[(is\_qexp~x_1);\ldots;(is\_qexp~x_n)]~[]~
    (atom\_~(is\_qexp}~\mathtt{E}))
\end{array}$$
then $\decode{\mathtt{\Gamma}}{\mathtt{E}}$ is defined and yields a
Proto-Quipper term $a$ such that $\Gamma \vdash \mathit{is\_qexp}~a$.
Additionally, $\encode{\Gamma}{\decode{\mathtt{\Gamma}}{\mathtt{E}}} =
\mathtt{E}$ and $\decode{\mathtt{\Gamma}}{\encode{\Gamma}{a}} = a$.
The above results correspond to Lemma 21 (called \textit{Validity of
  Representation}) and Lemma 22 (\textit{Completeness of
  Representation}), respectively,
in~\cite{FeltyMomigliano:JAR12}.\footnote{We also impose the
  restriction that the Coq derivation must be \textit{minimal} in the
  same sense as described there.  See~\cite{FeltyMomigliano:JAR12} for
  details.}

Proving the adequacy of the encoding of inference rules is similar.
Note that in the above statements, well-formedness was expressed as a
set of inference rules of a judgment of the form $\Gamma \vdash
\mathit{is\_qexp}~a$, and the proofs of these statements require
showing a bijection between proofs (on paper) using the inference
rules of this judgment, and proofs using our encodings of
$\mathtt{prog}$ and $\mathtt{seq}$.  The same approach is used to
prove the adequacy of the other two judgments $\Gamma;Q\vdash a:A$ and
$[C,a]\rightarrow [C',a']$.
Additionally, for all judgments except well-formedness, \emph{internal
  adequacy} lemmas must be proven.  Such lemmas correspond to Lemma
20, clauses 2 and 4 for the reduction and typing judgments,
respectively, of MiniML in~\cite{FeltyMomigliano:JAR12}.\footnote{A
  variety of other internal adequacy lemmas are shown
  in~\cite{FMP:JAR15}.}  They are called internal adequacy lemmas
because they can be formalized in Hybrid and they are an important
part of the general adequacy proofs for these judgments.
In general, such properties state that whenever an OL judgment that is
defined as part of the definition of $\mathtt{prog}$ can be proved,
then the OL terms in this judgment are well-formed.  The
$\mathtt{abstr}$ conditions are important for proving these lemmas.
The following theorem expresses internal adequacy for the
Proto-Quipper typing judgment.

\begin{flushleft}
\texttt{Lemma~hastype\_isterm\_ctx~:}\\
\texttt{~~forall~(M:qexp)~(T:qtp)~(iq~it~lt:list~atm),}\\
\texttt{~~ctxR~iq~it~lt~-{>}}\\
\texttt{~~seq\_~it~lt~(atom\_~(typeof~M~T))~-{>}}\\
\texttt{~~seq\_~iq~[]~(atom\_~(is\_qexp~M)).}\\
\end{flushleft}

\noindent
Note that it uses a context relation $\mathtt{ctxR}$ relating the
intuitionistic and linear contexts of the $\mathtt{typeof}$ predicate
to the intuitionistic context of the $\mathtt{is\_qexp}$ predicate.
(The linear context is always empty when proving well-formedness of
terms.)
The definition of this relation is as follows:

\begin{flushleft}
\texttt{Inductive~ctxR:~list~atm~-{>}~list~atm~-{>}~list~atm~-{>}~Prop~:=}\\
\texttt{|~nil\_cr:~ctxR~nil~nil~nil}\\
\texttt{|~cons\_q\_cr:~forall~(iq~it~lt:list~atm)~(x:qexp),~}\\
\texttt{~~proper~x~-{>}~ctxR~iq~it~lt~-{>}}\\
\texttt{~~ctxR~(is\_qexp~x::iq)~(is\_qexp~x::it)~(lt)}\\
\texttt{|~cons\_l\_cr:~forall~(iq~it~lt:list~atm)~(x:qexp)}\\
\texttt{~~(T:qtp),~~proper~x~-{>}~ctxR~iq~it~lt~-{>}~}\\
\texttt{~~ctxR~(is\_qexp~x::iq)~(is\_qexp~x::it)~(typeof~x~T::lt)}\\
\texttt{|~cons\_i\_cr:~forall~(iq~it~lt:list~atm)~(x:qexp)}\\
\texttt{~~(T:qtp),~proper~x~-{>}~ctxR~iq~it~lt~-{>}}\\
\texttt{~~ctxR~(is\_qexp~x::iq)~(typeof~x~T::is\_qexp~x::it)~lt.}
\end{flushleft}

\noindent
In addition the following two lemmas are needed.  The first is a
standard lemma about intuitionistic contexts as found in the examples
in~\cite{FMP:JAR15}.  The second is unique to linear contexts, and
thus is new.

\begin{flushleft}
\texttt{Lemma~qexp\_strengthen\_weaken:}\\
\texttt{~~forall~(M:qexp)~(Phi1~Phi2:list~atm),}\\
\texttt{~~(forall~(M:qexp),}\\
\texttt{~~~~In~(is\_qexp~M)~Phi1~-{>}~In~(is\_qexp~M)~Phi2)~-{>}}\\
\texttt{~~seq\_~Phi1~[]~(atom\_~(is\_qexp~M))~-{>}}\\
\texttt{~~seq\_~Phi2~[]~(atom\_~(is\_qexp~M)).}\\
\texttt{}\\
\texttt{Theorem~ctxRconcat:~forall~iq~it~lt~lt1~lt2,}\\
\texttt{~~ctxR~iq~it~lt~-{>}~~lt~=~lt1++lt2~-{>}}\\
\texttt{~~exists~it1~it2,}\\
\texttt{~~~~(forall~a,~In~a~it~-{>}~In~a~it1)~/{\char'134}}\\
\texttt{~~~~(forall~a,~In~a~it~-{>}~In~a~it2)~~/{\char'134}}\\
\texttt{~~~~ctxR~iq~it1~lt1~/{\char'134}~ctxR~iq~it2~lt2.}\\
\end{flushleft}

Note that the conclusion of the $\mathtt{hastype\_isterm\_ctx}$ lemma
only involves the term $\texttt{M}$.  No similar conclusion is
required for $\texttt{T}$ because Proto-Quipper types are defined
directly as an inductive type in Coq.  There are no binders in types,
and thus the correspondence between terms of type $\texttt{T}$ and
types of Proto-Quipper is direct.

We have proved a similar result for the reduction rules, showing that
the language's operational semantics does not lead to invalid
expressions (see~\cite{ourformal}).

\section{Conclusion}
\label{sec:concl}

We have presented our formalization of Proto-Quipper in Hybrid.  This
work involved encoding a linear specification logic and carrying out a
large case study in Hybrid, in the sense that we encode and reason
about the complete Proto-Quipper specification, the most complex OL
considered so far,
and we prove type soundness, one of the central results
in~\cite{RossPhD15}.

In order for Hybrid with a linear SL to become a fully operational
logical framework, we will need to provide suitable automation of
proofs, based on lessons learned from this case study.
Adding such automation is an important direction for future work.

Another important direction is extending the formalization to other
more complex properties and to other quantum programming languages.
For example, the other main result~\cite{RossPhD15} is a progress
theorem for Proto-Quipper, which is an obvious next step for us.  We
do not foresee any difficulty, though its proof will likely be as long
and detailed as the proof of type soundness. Also, it should be
straightforward to adapt the existing formal proofs to new versions of
Proto-Quipper.  For example, a new version called Proto-Quipper-M is
introduced in~\cite{RiosSelinger:QPL17}.  This version does not have
subtyping, which should mean that certain aspects of formal proofs of
its metatheory will be easier.  In the future, we hope to use our
system as an environment in which new Proto-Quipper metatheory can be
simultaneously developed and formalized as versions of the language
evolve.

It should also be possible to directly extend our work
from Proto-Quipper, which is fairly expressive though not Turing
complete, to the full Turing-complete Quipper language.  Similarly, we
expect to be able to extend this work directly in order to apply it to
other Turing-complete quantum lambda calculi such as those introduced
by Zorski et.\ al.~\cite{LagoMZ:MSCS09,LagoMZ:TCS10,Zorzi:MSCS16} and
Grattage et.\ al.~\cite{AltenkirchGrattage:LICS05,Grattage:ENTCS11}.
As another example, the QWIRE language in \cite{RandEtAl:QPL17} is
implemented in Coq and is expressive enough to include languages like
Proto-Quipper. The work in that paper focuses on proving properties of
quantum programs and program transformations, such as proving that a
program meets its formal specification.  In our framework, it would be
interesting to also study the meta-theory of this language.

A variety of other logical frameworks implementing linear logic have
been developed.  The ordered linear logic (OLF) mentioned
earlier~\cite{Polakow01phd} is one such example, where type soundness
for a \emph{continuation}-based abstract machine for the functional
programming language Mini-ML (an OL that is simpler than
Proto-Quipper) is proven~\cite{FeltyMomigliano:JAR12} following the
statement in~\cite{CervesatoP02}.  Another example of OLs that benefit
from a framework based on linear logic are those with imperative
features.  A common thread of such examples is that they contain the
notion of updatable state, which can be handled fairly directly by the
linear features of the framework.  The case study we present in this
paper is the first one in Hybrid where the OL itself is linear, in the
sense that the linear lambda calculus forms the core of Proto-Quipper.

Examples involving mutable state have motivated a variety of proposals
for frameworks based on linear logics that support HOAS.  For other
examples that benefit from linear features, see the overview
in~\cite{miller04llcs}.  Examples of frameworks proposed include
Lolli~\cite{Lolli}, Forum~\cite{Forum}, and LLF~\cite{CervesatoP02}.
The logical framework LF~\cite{Harper93jacm} and its implementation in
Twelf represents one of the earliest logical frameworks supporting
HOAS and based on minimal intuitionistic logic.  LLF is a conservative
extension of LF with multiplicative implication, additive conjunction,
and unit.

Type soundness proofs for various OLs has been a common benchmark for
logical frameworks supporting HOAS and implementing an intuitionistic
logic, starting with some of the earliest logical frameworks like
LF~\cite{TwelfSTLC}. In~\cite{MartinPhD2010}, Mini-ML with mutable
references, an imperative version of the Mini-ML language mentioned
above, is studied in Hybrid.  Five versions of type soundness are
proved using three different SLs, intuitionistic, linear, and OLF as
implemented in~\cite{FeltyMomigliano:JAR12}, and their formalizations
compared.

\section*{Acknowledgements}

The authors would like to thank Julien Ross and Peter Selinger for
useful discussions on technical details as well as on approaches and
directions for this work.  We would also like to thank the reviewers
for useful comments for improving this paper.


\end{document}